\documentclass[article]{aa}

\usepackage{float}
\usepackage{graphicx}
\usepackage{placeins}
\usepackage[normalem]{ulem}
\usepackage{txfonts}
\usepackage{hyperref}
\usepackage[dvipsnames]{xcolor} 
\hypersetup{
    colorlinks,
    citecolor=RoyalBlue,
    filecolor=RoyalBlue,
    linkcolor=RoyalBlue,
    urlcolor=RoyalBlue
}

\title{When bars and spirals conspire: Recurrent build-up of the nuclear regions of disc galaxies}
\titlerunning{When bars and spirals conspire}

\author{T. Boin\inst{1}\and
      S. Khoperskov\inst{2}\and
      P. Di Matteo\inst{1}\and
      N. Hoyer\inst{1}\and
      A. Mastrobuono-Battisti\inst{3,4,5}\and
      N. Ryde\inst{6}\and
      F. Combes\inst{7,8}\and
      M. Schultheis\inst{9}\and
      M. Haywood\inst{1}
}

\institute{
LIRA, Observatoire de Paris, Université PSL, Sorbonne Université, Université Paris Cité, CY Cergy Paris Université, CNRS, 92190 Meudon, France\\\email{tristan.boin@obspm.fr}
\and Leibniz-Institut für Astrophysik Potsdam (AIP), An der Sternwarte 16, 14482 Potsdam, Germany
\and Dipartimento di Fisica e Astronomia “Galileo Galilei”, Università di Padova, Vicolo dell’Osservatorio 3, 35122 Padova, Italy
\and Istituto Nazionale di Astrofisica - Osservatorio Astronomico di Padova, Vicolo dell’Osservatorio 5, 35122 Padova, Italy
\and Dipartimento di Tecnica e Gestione dei Sistemi Industriali, Università degli Studi di Padova, Stradella S. Nicola 3, 36100 Vicenza,Italy
\and Division of Astrophysics, Department of Physics, Lund University, Box 118, 221 00 Lund, Sweden
\and LUX, Observatoire de Paris, PSL Research University, Sorbonne Université, CNRS, 75014 Paris, France
\and Collège de France, 11 place Marcelin Berthelot, 75231 Paris, France
\and Université Côte d’Azur, Observatoire de la Côte d’Azur, Laboratoire Lagrange, CNRS, Blvd de l’Observatoire, 06304 Nice,
France
}

\date{Received XXX; accepted XXX}

\abstract{
The assembly history of the central regions of disc galaxies is regulated by dynamical processes that trigger gas infall events and lead to active star formation in nuclear stellar discs and in nuclear stellar clusters. In the Milky Way, recent studies of its nuclear regions have revealed a complex star formation history, with an initial burst associated with the formation of the Galactic bar followed by a non-constant star formation rate.
}{
In this work, we study the formation and evolution of nuclear structures as well as the link between the formation of large-scale structures and the nuclear stellar disc and star cluster. Our goal is to investigate the effects of the bar and spiral arms on the dynamics of the gas and, as a result, on the star formation history of nuclear stellar structures.
}{
To this aim, we ran and analysed a simulation of an isolated Milky Way-like galaxy with the \textit{SWIFT} \textit{N}-body plus hydrodynamics simulation code, including star formation and stellar feedback from supernovae types Ia and II. We start from a live dark matter halo and a pre-existing stellar and gaseous disc with a 20\% gas fraction, which quickly forms a bar, a boxy-peanut bulge, spiral arms, and nuclear structures. We studied the star formation history of these regions and how they relate to variations in the galactic bar length, strength, and pattern speed. We also investigated the role of spiral arms and their interaction with the bar.
}{
We find that the star formation history of the nuclear regions displays a main burst at the time of the formation of the bar due to bar-driven gas inflows. After bar formation, we found secondary regularly spaced star formation bursts that do not appear in the disc star formation history. These secondary bursts occur when the spiral arms and the bar -- which rotate at different pattern speeds -- reconnect, triggering secondary gas inflow events.
}
{
The interaction of spiral arms and the galactic bar can enhance non-axisymmetric features in the disc, and thus trigger bar-driven gas infall even after the bar has formed. Our simulation suggests that these bar-spiral reconnection events are imprinted in the star formation history of the nuclear stellar clusters and nuclear stellar discs as episodic star formation bursts.
}

\begin{document}

\maketitle
\nolinenumbers

\section{Introduction}
About two-thirds of disc galaxies in the local Universe have been shown to host a bar~\citep{2000AJ....119..536E, 2007ApJ...657..790M, 2008ApJ...675.1194B, 2011MNRAS.411.2026M,Erwin_18}, and this fraction decreases with increasing redshift~\citep{2008ApJ...675.1141S, 2010MNRAS.409..346C, 2014MNRAS.438.2882M,LeConte_24}. Bars are elongated structures mostly made of stars that rotate around the centre of a galaxy with a pattern speed that can be assumed constant at first approximation but in fact may vary with time because of angular momentum redistribution with{in the galaxy itself driven by internal or external processes \citep{2013MNRAS.429.1949A, 2014MNRAS.438L..81A, 2017MNRAS.464.1502M, 2020A&A...642L..12L, 2021MNRAS.502.3085G, 2025ApJ...991L..52L, 2026ApJ..1000..161C, 2026MNRAS.547ag371M}. Stellar bars can also thicken with time, forming central out-of-the-plane structures that do not have a classical spheroidal shape but rather so-called boxy-peanut (hereafter B/P) morphologies~\citep{Combes_93,BP_coeff}. The B/P bulges are common in disc galaxies. Indeed, nearly half of edge-on galaxies with bulges have such a shape~\citep{2016ASSL..418...77L, 2017MNRAS.468.2058E}, which can be acquired by a disc galaxy either through a vigorous bulge-formation event \citep{1991Natur.352..411R, 2006ApJ...637..214M} or through a smooth trapping of stars at the vertical resonance with the bar~\citep{1990A&A...233...82C, 2014MNRAS.437.1284Q, 2020MNRAS.495.3175S, 2025A&A...699A.234L}. Our Galaxy, the Milky Way, is a typical disc galaxy in this respect, as it features both a central bar~\citep{1991ApJ...379..631B, 2005ApJ...630L.149B,Wegg_15} and a B/P structure~\citep{2010ApJ...724.1491M, 2010ApJ...721L..28N, 2013MNRAS.435.1874W, 2016AJ....152...14N}. The link between the Galactic bar and B/P structure, and more generally between bars and B/P bulges, has been extensively explored in the past. These structures are now understood in the more general context of the
formation and evolution of galaxy discs and their stellar populations, out of which bars -- and
some time after B/P bulges -- form, driven by secular or environmental processes \citep{2019A&A...624A..37L,2024A&A...683A.196G, 2025MNRAS.540.2031L}. \\  Whilst in disc galaxies with masses similar to that of the Milky Way, bars and B/P bulges typically have dimensions of a few kiloparsecs \citep[in the Milky Way, the semi-major axis of the bar and bulge are estimated to be, respectively, of 4--5 kpc, and about 2 kpc; see ][]{2016ARA&A..54..529B} an investigation of their central regions revealed the presence of additional complex stellar structures, such as nuclear star clusters (NSCs; \citealt{2002AJ....123.1389B, 2006ApJS..165...57C,Georgiev_16, Neumayer_20}) and, often, nuclear stellar discs (NSDs; \citealt{2002ApJ...573..131P, 2010MNRAS.407..969L, 2020A&A...643A..14G}). 

Nuclear star clusters appear as dense and massive compact stellar systems located at the dynamical centre of their host galaxy~\citep{Neumayer_12,Neumayer_20}. In our Galaxy, the NSC has an effective radius of about 4 - 7 pc~\citep{2014A&A...566A..47S, 2016ApJ...821...44F, 2020A&A...634A..71G}, and it contains several million stars, for a total stellar mass of a few $10^7~\rm M_\odot$.  The NSC is the densest stellar structure of the Galaxy, and it is denser than any known Galactic globular cluster, and whilst its stellar populations are generally old \citep[older than 5 Gyr, see][]{Schodel_20, Neumayer_20}, its very centre also contains a minor population of young stars \citep{Schodel_20, 2023ApJ...944...79C, Nogueras_Lara_20, 2026A&A...708A..77G}. 

Different physical mechanisms have been proposed to explain the formation of NSCs in galaxies: orbital decay and merging of globular clusters in the central regions of galaxies \citep{1975ApJ...196..407T, 1993ApJ...415..616C, 2008ApJ...681.1136C, 2008MNRAS.388L..69C}, in situ star formation driven by magnetorotational instability \citep{2004ApJ...605L..13M} or by a combination of large-scale gravitational torques that funnel gas into the central regions, and magnetic torques and cloud-cloud collisions that act at smaller scales \citep{1990Natur.345..679S}. In situ star formation may be recurrent and regulated by supernovae explosions, stellar winds, and the dissipation of turbulent energy  \citep{1982A&A...105..342L,Bonnell_08,Hobbs_09,Mapelli_12,Trani_18,Goicoechea_18}. The merging of globular clusters in the inner regions of galaxies may account for the presence of old stellar populations in NSCs and may also explain some of the properties of the MW NSC, such as its density profile \citep{2012ApJ...750..111A}, rotation, and flattening \citep{2017MNRAS.464.3720T}. However, this scenario alone cannot explain the presence of young as well as metal-rich \citep{Feldmeier_krause_20} stars found in many NSCs, including in our own Milky Way \citep{Schultheis_21}. Abundance studies of the NSD by \cite{Ryde_25} and of the NSC by \cite{Nandakumar_25} also support a different formation scenario for the nuclear regions. These findings indeed require some amount of in situ star formation in which gas can accumulate in the nuclear regions through different mechanisms (bar-driven infall, galaxy mergers, tidal compression, etc.) and then form stars, as in the case of NGC 6946, which displays a nuclear starburst, a 'bar within bar' configuration, and inner structures that \cite{Romeo_15} associate with gravitational instability and radial inflow. It has also been suggested that these different mechanisms may dominate at different galaxy masses, with globular cluster infall being the dominant process at play in galaxies with stellar masses below $\sim 10^9~\rm M_\odot$ and in situ formation becoming the dominant physical mechanism responsible for the formation of NSCs in more massive galaxies \citep{Neumayer_20,Fahrion_21}. In their 'hybrid' scenario, \cite{Guillard_16} formed an NSC by star cluster infall, which retained some amount of gas, triggering recurrent in situ star formation. Finally, observations of NSC spectra in late-type galaxies support multiple star formation episodes, leading to a range of stellar populations \citep{Walcher_06,Rossa_06,Carson_15}.

Nuclear star clusters are often surrounded by flattened, rotating stellar discs -- the so-called NSDs -- which are found both in early- and late-type galaxies \citep{Schultheis_25}. In the Milky Way, the NSD  has an exponential radius of about 100 pc and a mass of a few $10^9~\rm M_\odot$, and it dominates the Galactic potential for a few hundred parsecs \citep{Sormani_22,Schultheis_25}. Nuclear stellar discs  -- thanks to their flattened morphology and rotation-driven dynamics -- are thought to result from in situ star formation, which arises from the transformation of gas within the central hundreds of parsecs of a galaxy into new stars. This obviously requires the presence of gas in these regions for several billion years, a timescale consistent with the age spread typically observed for the stars and stellar populations of NSDs \citep{bittner20, pessa23}. Using Jeans modelling \citep{Sormani_20} or self-consistent equilibrium models \citep{Sormani_22}, the properties of the MW NSD have been constrained, with the authors finding properties consistent with an 'inside-out' formation scenario. This is in line with the findings of \cite{bittner20} in 21 nuclear discs of galaxies within the TIMER survey \citep{Gadotti_19}. Extensive observational work has also been done to map the stellar populations of the MW NSD and its star formation history \citep[SFH;][]{Nogueras_Lara_20,Nogueras_Lara_21,Nogueras_Lara_22b,Nogueras_lara_23b,Schodel_23,Nogueras_lara_24a}.

In disc galaxies with a stellar bar, the bar itself can act as the way through which gas can be transported from regions extending over several kiloparsecs towards the central regions. Stellar bars are in fact asymmetric mass distributions, and as such, they exert negative torques on the gas within the galaxy disc, particularly in the regions inside the bar’s corotation~\citep{Combes_1985, Athanassoula_92, Shlosman_89}. As a result of these torques, the gas loses some of its angular momentum until it reaches the central regions, where it can form a disc whose radial extent -- as well as that of the nuclear ring often associated with it -- may be linked to the location of the bar’s Lindblad internal resonance \citep{1992MNRAS.259..328A, Athanassoula_92, 1996ASPC...91..286C,Sormani_18,Sormani_24}. A recent study by \cite{Santiago_26} on the nearby spiral galaxy NGC 6946 revealed a recent starburst episode $\sim$ 0.6~Gyr ago caused by enhanced gas infall driven by a nested bar system, followed by a declining phase until today. Star formation in its central regions stays higher than in the outer regions due to the high amount of gas accumulated from bar-driven inflows.

\begin{table}
    \caption{Initial parameters for each of the components of the $N$-body simulation used in this work.}\label{tab:model}
    \begin{tabular}{c | c c c c c }
        Component & M  & r & $h_z$ & $n_p$& $\sigma_{r,0}$\\ 
         &  $[10^{10}\ \rm{M}_\odot]$ & [kpc] &  [kpc] & $[10^6]$ & [km/s]\\ 
        \hline
       D1 & $3.8 $ & 4.8 & 0.15 & 4 & 100\\  
       D2 & $1.9$ & 2   & 0.3  & 2.5 & 120\\
       D3 & $1.5$ & 2   & 0.6  & 2 & 150\\
       DG & $1.8$ & 10   & 0.05  & 0.4 & 10\\
       DM & $13.3$ & 10  & -    & 5 & -
    \end{tabular}
    \tablefoot{From left to right: Total mass, characteristic radius, scale height, number of particles, and central radial velocity dispersion. The D1, D2, D3, DG, and DM components respectively correspond to the stellar thin, intermediate, and thick discs; the gas disc; and the dark matter halo. The mass of the dark matter halo is the mass contained within 100~kpc.}
\end{table}

The most significant inflows of gas into the central regions of a barred galaxy typically occur at the time the bar forms~\citep{Baba_20,Verwilghen_24}. It is in fact at this time that the gas, initially assumed to be in nearly circular orbits, undergoes the most significant torques. Therefore, dating the oldest stars present in the NSD of a barred galaxy can be a way of dating the time when the bar in that galaxy formed. Using this approach, it has been suggested that the formation of our Galaxy’s bar occurred around 8 billion years ago~\citep{Sanders_24}, an estimate that is consistent with that obtained by dating the epoch at which the strongest migration of old metal-rich stars from the central regions – triggered by the formation of the bar itself – took place~\citep{2024A&A...690A.147H}. This same approach -- dating the epoch of formation of bars by dating NSDs – has also been applied to external galaxies, resulting in a variety of bar formation epochs for nearby galaxies (see~\citealt{De_Sa_Freitas_23}). 
  
Bars are not the only asymmetrical structures found in disc galaxies. They are often accompanied by spiral arms, which extend from the regions outside the bar to the edges of the discs. Spiral arms generally have multiple modes~\citep{1977PNAS...74.4726B, Sellwood_14,Sellwood_19}, can be long-lived~\citep{Quillen_11,Minchev_12}, and have different pattern speeds from those of the bar itself~\citep{ 1999A&A...348..737R, Sellwood_88}. This implies that, on a recurring basis, spiral arms reconnect and disconnect from the bar ends~\citep{Tagger_87,Masset_97}. This regular overlap between a bar and the system of spiral arms can affect the estimation of the bar’s parameters, such as its length and amplitude~\citep{Hilmi_20}, as well as the disc radial velocity field~\citep{Vislosky_24}. When the spiral arms reconnect with the bar, indeed, the latter appears longer and stronger than when the two are disconnected. Bar-spiral arm reconnection has also been shown to generate bursts in both radial migration and star formation at the bar ends, with a significant increase in the local star formation~\citep{Marques_25}.
  
Given the context outlined above, in this article, we continue the study of the formation and evolution of the central regions of disc galaxies. In particular, we present and analyse a hydro-$N$-body simulation of a Milky Way-type galaxy, and we demonstrate how the presence of a bar and spiral arms and their regular overlap can lead to recurrent bursts of star formation in the nuclear regions of the simulated galaxy, driven by the torques exerted by a large portion of the disc, at the time these reconnections occur. As we show, together with the first strong star formation burst occurring at the time of bar formation, these regular secondary bursts can also contribute to the mass growth of both an NSC and an NSD, resulting in an age spread of their populations.

The paper is organized as follows: In Sect.~\ref{sec:num_mod} we introduce our simulation setup. In Sect.~\ref{sec:results} we describe the large- and small-scale structures that form in the galaxy, namely a bar, a B/P bulge, an NSD, a nuclear bar, and a nuclear stellar cluster. We analyse the time evolution of their properties as well as their star formation rates (SFRs), and we link them to bar-spiral interactions. In Sect.~\ref{sec:discu} we discuss our results and compare them to observations of the Milky Way nuclear regions as well as results from other numerical studies. Finally, we summarise our findings in Sect.~\ref{sec:conclu}.

\section{Numerical modelling}\label{sec:num_mod}

In this work, we analyse an $N$-body/hydrodynamics simulation of an isolated Milky Way-like galaxy with the following composition:
\begin{itemize}
    \item  A dark matter halo of mass $1.33 \times 10^{11}~\rm{M}_\odot$ within 100~kpc and initially following a Hernquist profile \citep{Hernquist_90} with a scale radius 10~kpc.
    \item  A gaseous disc of mass $1.8 \times 10^{10}~\rm{M}_\odot$ following an exponential profile.
    \item  A pre-existing composite stellar disc made up of three components following exponential profiles, namely a thin, intermediate and a thick disc, having respectively shorter scale heights, higher scale lengths, and lower velocity dispersions.
\end{itemize}

Masses, characteristic radii, scale heights, number of particles and central radial velocity dispersions of each component are summarised in Table~\ref{tab:model}.

The initial conditions (positions and velocities) for such a configuration were generated with the AGAMA code \citep{AGAMA}, in which the components are constructed through an iterative process in a self-consistent manner, leading to the rotation curve reported in Appendix~\ref{app:rotation_curve}. 
In particular, the properties of the composite stellar disc are based on the simulation originally introduced in \cite{Fragkoudi_17}, albeit with exponential discs instead of Miyamoto-Nagai ones, and further used in \cite{Fragkoudi_18,Khoperskov_18,Fragkoudi_19,Boin_24}. Such simulations reproduce several observed properties of the Milky Way bulge, including longitudinal and latitudinal metallicity gradients, metallicity distributions and their variations,  its chemo-kinematic relations and a differential mapping of stellar populations in the inner bulge region based on their initial velocity dispersion.

The simulation was performed using the SWIFT $N$-Body/Smoothed Particle Hydrodynamics (SPH) simulation code \citep{SWIFT}. Gravitational forces are computed through a fast multipole method; the gravitational softening length is set to a constant 50 pc for all particles, and we used an opening angle of $\theta=0.7$ for the gravity computation. For the gas, the SPHENIX smoothed-particle hydrodynamics scheme \citep{sphenix} is used, with a CFL condition of 0.1 and a minimal smoothing length of 5~pc. A star formation prescription is included and follows a Schmidt law with a star formation efficiency of $\epsilon_{eff}=0.01$.  The subgrid model is based on the EAGLE model \citep{EAGLE_1,EAGLE_2}. The simulation includes stellar feedback events in the form of SNIa, SNII, and AGB stars. The energy released per SNIa and SNII event is set to the canonical value of $10^{51}$ erg, with the SNII energy release following a scaling law depending on metallicity and local gas density (see Eq.~7 in \citealt{EAGLE_1}), with minimal and maximal energy fractions of $f_{th,min}=0.0388$ and $f_{th,max}=1$, which are used in order to take into account the increase of thermal loss in the ISM due to metals in the gas. The integration is resolved with a variable time-step, calculated based on the different criteria imposed by the current state of the simulation, i.e. the gravitational accelerations and the Courant condition for the hydrodynamics. The Courant condition provides an upper bound on the simulation timestep based on the gas' velocity and spatial resolution to ensure its dynamics are correctly integrated. Outputs are saved every 10~Myr.

\section{Results}\label{sec:results}
\subsection{Large-scale and nuclear structures}

\begin{figure*}[!htbp]
    \centering
    \includegraphics[width=\linewidth]{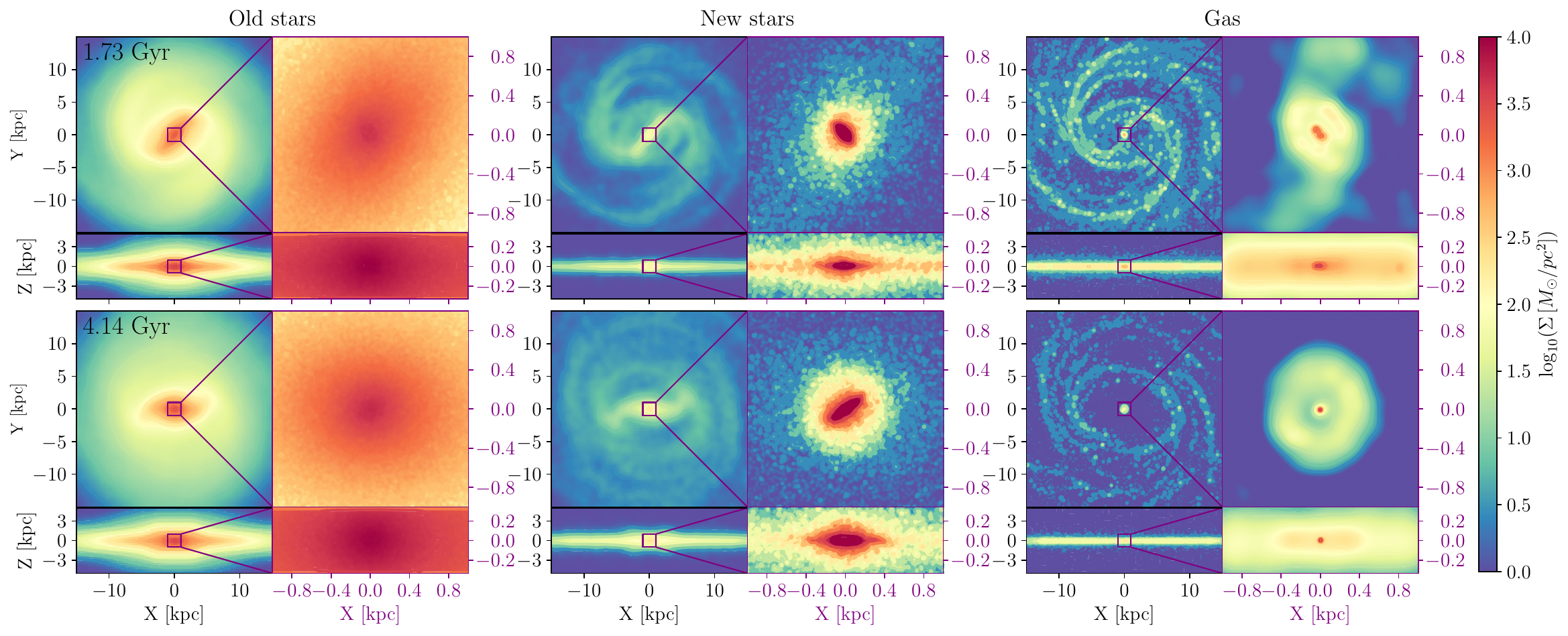}
    \caption{Density face-on maps of the simulation at rescaled times of 1.73 Gyr, after the bar formation (top rows), and 4.14 Gyr, at the end of the simulation (bottom rows), for old stars (left panels), new stars (middle panels), and gas (right panels). The odd column panels show maps at the galactic scale (30~kpc-wide), while the even column panels (highlighted in purple) show maps in the inner 1~kpc region, restricted to particles within $\rm |z| < 300~pc$.
    }
    \label{fig:face_on_maps}
\end{figure*}

In this section we give an overview of the different structures that form in our simulation, their properties and formation time, as well as their further evolution, as illustrated at two snapshots (t=2.93~Gyr and t=7.00~Gyr, the end of the simulation for unscaled times, see below) in Fig.\ref{fig:face_on_maps}.  The simulation is stopped at 7~Gyr as it has then reached a steady state in terms of structure formation. From here on, the simulation outputs are rescaled to the present-day Milky Way by requiring that the semi-major axis of the bar at the end of the simulation match the observed Galactic bar length. Following \cite{Wegg_15}, we adopt a Milky Way bar length of $5.0 \pm 0.2\rm~kpc$. The unscaled bar length measured at t=7~Gyr is 8.5~kpc, giving a rescaling factor of $\lambda=\frac{5}{8.5}\sim0.6$. The consistency of the simulation is kept by keeping velocities untouched and rescaling masses and time according to the virial theorem $v^2\sim\frac{GM}{R}$, i.e. $R\xrightarrow{}\lambda R$, $v\xrightarrow{}v$, $M\xrightarrow{}\lambda M$ and $t\xrightarrow{}\lambda t$. This particular choice of rescaling ensures dynamical consistency of the simulation by ensuring that orbital quantities obey the same laws independently of the scaling chosen. Hence, the final snapshot of the simulation reaches $t=~4.14~\rm Gyr$, and the total stellar mass is $4.93\times 10^{10}\rm~M_\odot$, close to the low end of the $6.08\pm 1.14\times 10^{10}\rm~M_\odot$ estimate given by \cite{Licquia_15} for the Milky Way.

\subsubsection{Bar, B/P bulge, and spiral arms}\label{sec:bar_bulge_spiral}
Starting from the pre-existing stellar discs and gaseous disc, a bar forms around $t=$1~Gyr with respect to the start of our simulation. This bar formation time corresponds to the maximal value of the bar strength, i.e. of  the ratio $\rm A_2/A_0$ as a function of radius shown in the inset top panel of Fig.~\ref{fig:bar_properties}, and with Fourier coefficients defined as

$$
a_m = \sum_{i=0}^N\cos(m\theta_i),
$$
$$
b_m = \sum_{i=0}^N\sin(m\theta_i),
$$
$$
A_i = \sqrt{a_m^2+b_m^2}.
$$
The zeroth term measures the axisymmetric component of the distribution, and the second term measures the bisymmetry of the non-axisymmetric part of the distribution.
The bar can be seen in the inset face-on maps in Fig.~\ref{fig:face_on_maps}, in the pre-existing stars, as well as the newly born stars. Shortly after its formation, the bar buckles to develop a B/P bulge, whose strength we measure by the normalised C$_{\rm 2z,max}$ coefficient, as defined in \cite{BP_coeff}, and reported in the bottom panel of Fig.~\ref{fig:bar_properties}. Along with the formation of a galactic bar, spiral arms develop over time, as can be seen in the inset face-on maps of Fig.~\ref{fig:face_on_maps}, more strongly traced by new stars and gas (middle and right inset columns).

The global properties of the bar, i.e. its strength, length and pattern speed and their temporal evolution, are reported in Fig.~\ref{fig:bar_properties}. The bar length $\rm R_{bar}$ is given by the radius at which the bar strength $\frac{A_2}{A_0}$ drops below 80\% of its maximal value. Globally, the bar strength is slightly increasing from bar formation at t=1~Gyr to the end of the simulation, while the bar length fluctuates around an approximately constant value and the bar pattern speed decreases with time. After the bar formation at t=1~Gyr, a periodic signal is observed in both the bar pattern speed, strength and length, the three being correlated. The period of the fluctuation increases from 0.29~Gyr at t=1.5~Gyr to 0.55~Gyr at t=4.14~Gyr. To verify the robustness of this bar length measuring technique, we compare it in Appendix~\ref{app:bar_length} to two other methods, which yield similar periodic fluctuations. In the third panel of Fig.~\ref{fig:bar_properties}, we report the strength of the B/P bulge, whose formation time we place at 1.25~Gyr, the time at which $\rm C_{2z,max}$\footnote{We define $\rm C_{2z,max}$ as $\frac{|\sum_{j=1}^{N_*} z_j*e^{i2z_j/(5z_0)}|}{\sum_{j=1}^{N_*}|z_j|}$ to obtain a normalised measure of the B/P strength.} is maximal (see the peak of $\rm C_{2z,max}$ in the third panel of Fig.~\ref{fig:bar_properties}). The bar pattern speed, which we measure through the method described in \cite{Dehnen_23} and based on the Tremaine-Weinberg approach \citep{Tremaine_84}, decreases from 34.5$\pm0.4$~km/s/kpc at t=0.6~Gyr to 22.8$\pm0.1$~km/s/kpc at t=4.14~Gyr (rescaled from its initial value of 13.7~km/s/kpc at t=7~Gyr), which we note is lower than current estimates of the MW bar pattern speed. For example, \cite{Portail_17} found $\rm\Omega_b=39.0\pm3.5~km/s/kpc$, \cite{Sanders_19} found $\rm\Omega_b=41\pm3~km/s/kpc$, and \cite{Clarke_19} found $\rm\Omega_b=37.5~km/s/kpc$. On top of this decrease in the rotation rate of the bar, the pattern speed oscillates with amplitudes of $\sim$2~km/s/kpc, fluctuations which are correlated with the bar length and strength. While the global temporal evolution of the bar/bulge properties follows well-known trends (e.g. the global slowing down of the bar is expected through angular momentum transfer with the dark matter halo, the bar strength build-up is exponential at bar formation time), the periodic fluctuations on top of these trends are the main object of our analysis.

\subsubsection{Nuclear star cluster and nuclear stellar disc}

At the sub-kiloparsec scale (which we refer to as the nuclear region), multiple structures form, as can be seen in the panel of Fig.~\ref{fig:face_on_maps}. A disc-like structure forms in the central 500~pc shortly after the main galactic bar formation ($\sim$1.25~Gyr), within which a nuclear bar-like structure appears, as can be seen in the final snapshot (lower middle panel). These structures only faintly appear in the old~(pre-existing) star maps, as they are in the majority composed of stars born from the initial gaseous disc. The gas maps display a cluster-like structure in the inner 50 parsecs, and a ring-like structure of radius $\sim$ 200~pc at its creation at t$\sim$1.25~Gyr, and whose size increases to 400~pc by t=4.14~Gyr (right columns of Fig.~\ref{fig:face_on_maps}).

The disc-like structure, which we refer to as an NSD, has a mean $v/\sigma=1.5$ as measured from stars with radius 150~pc < R < 500~pc and heights |z| < 200~pc, in order to isolate the NSD from the cluster-like structure. In Appendix~\ref{app:v_over_sigma} we plot the profiles of azimuthal velocity, total velocity dispersion and $v/\sigma$. The latter we refer to as an NSC, and has a mean $v/\sigma=0.5$, as measured from stars with radius R < 150~pc and heights |z| < 200~pc. We note that this delimitation is somewhat arbitrary and does not reflect a sharp boundary between the two structures, but was rather chosen for our analysis as it corresponds to two delimited regions of star formation, as seen in the third panel of Fig.~\ref{fig:v_over_sigma}. The $v/\sigma$ profile of the NSD peaks at R=380~pc, with a value of 2.3, while the NSC displays a monotonously increasing $v/\sigma$ with radius, with a vanishing value at the centre, consistent with the $|V|/\sigma$ profiles of galactic nuclei reported by \cite{Pinna_21}. This is consistent with an NSD that is rotationally supported and an NSC that is pressure-supported. For the NSD, we measure a mean velocity dispersion of $\langle\sigma\rangle\simeq 65$ km/s, and a mean azimuthal velocity of $\langle v_\phi\rangle\simeq101$ km/s, in the range of observations of the kinematics of the MW NSD \citep{Schonrich_15,Schultheis_21,Shahzamanian_22,Sormani_22}. For the NSC, the mean velocity dispersion is $\langle\sigma\rangle\simeq 64$ km/s and the mean azimuthal velocity $\langle v_\phi\rangle\simeq33$ km/s. In their review of NSCs, \cite{Neumayer_20} reports $\sigma=60\pm5$ km/s and $v_{los}=50\pm3$ km/s for the Milky Way NSC, at its effective radius $r_{eff}=4.2$ pc, giving a $v/\sigma$ value of 0.83. In their one-population dynamical model, \cite{Feldmeier_25} find a $\rm V_{LOS}$ of 41~km/s and a $\rm \sigma_{LOS}$ of 58~km/s at 5~pc, leading to a $v/\sigma$ of 0.71, both slightly higher than our value.

\subsection{Star formation in the disc and in the nuclear regions and its recurrence}

Some of the structures we described in the previous section -- such as the NSD and the nuclear bar -- consist almost exclusively of stars that formed, over time, from the initial gaseous disc. It is thus natural to quantify the SFH of the simulated galaxy, both on the large and on the nuclear scale.

In the top panel of Fig.~\ref{fig:SFH}, we separate the SFR for stars with $\rm R_{birth}<1~kpc$ (turquoise curve) and for all stars (purple curve). The solid curves correspond to the measured SFR, as measured by the gas mass turned into stars within a time bin, while the dashed curves correspond to the present-day SFR, i.e. the age distribution of stars, weighted by their mass at t=4.14~Gyr. The difference between the two curves arises from stellar evolution, with massive stars losing mass over their lifetime. Thus, the latter corresponds to what an observed age distribution would look like to us today, whereas the former is the real SFR, weighted by the masses of stars at their birth. The two SFRs display the same evolution, with the main difference arising from the fact that the former is higher at earlier times. We include both in Fig.~\ref{fig:SFH}, as the latter is directly comparable to observations, whereas the former reveals the true amplitude of fluctuations in star formation. In the following, SFR refers to the real SFR; however, the conclusions are qualitatively the same for the present-day SFR. The SFR in the outer regions is the highest at the very beginning of the simulation, and progressively decreases with time, as we simulate an isolated galaxy, without possible gas additions from the accretion of filamentary material. A burst of star formation occurs at the time of bar formation (marked as a vertical dashed line at t$\sim$1~Gyr), and lasts around 0.5~Gyr. The subsequent temporal evolution of the SFR is overall decreasing, with slight periodic variations. Looking at the star formation in the nuclear regions (i.e. inner 1~kpc), we observe no star formation activity before the bar forms. It is only at the time of bar formation that an intense SF burst takes place, which happens at approximately the same time as the global SF burst, with a time lag of 70~Myr. As for the large-scale SFR, the nuclear SFR also decreases after bar formation. In the nuclear regions, however, the subsequent SF variations are significant: in the 2-4~Gyr range, the background SFR is of around 0.2~$\rm M_\odot/yr$, whereas the bursts -- each lasting on average 200~Myr --  increase it to around 0.6~$\rm M_\odot/yr$.

The star formation burst linked to the bar formation arises due to the high angular momentum loss that this event induces on the gas, as the non-axisymmetric bar structure develops, which is reflected in the peak value of the $\rm \left(\frac{A_2}{A_0}\right)_{max}$ of the top panel of Fig.~\ref{fig:bar_properties} at t=1~Gyr. The peak of the BP strength seen in the third panel of the same figure, reached at t=1.25~Gyr, coincides with a burst of SF seen in the inner region SFH in Fig.~\ref{fig:SFH}.

In the bottom panel of Fig.~\ref{fig:SFH}, we plot the cumulative mass gained in the central 1~kpc region. We divide the time evolution into three phases: a first quiescent phase before bar formation, followed by an SFR burst linked to bar formation, and ending with a more quiescent but not fully quenched phase. The second phase, which we define as the time range between 0.75 and 1.25~Gyr, builds up a total of $0.83\times10^9\rm~M_\odot$, with an average SFR of $1.65\rm~M_\odot/yr$. The SFR over the third phase is lower, with a mean value of $0.34\rm~M_\odot/yr$; however, it spans close to 3~Gyr from 1.25 to 4.14~Gyr. The secondary star formation peaks contribute to maintaining star formation activity, resulting in a mass build-up of $10^9\rm~M_\odot$ over this time range. Thus, we find that the inner regions are capable of forming as many, if not more, stars after the formation of the bar than during it. The star formation is declining but has not yet ceased entirely, and is regularly reignited by SF bursts. In our simulation, these latter account for about 30-40\% of the total mass formed over this phase, as measured from the time-integrated SFR after the bar formation, from t=1.5~Gyr to t=4.14~Gyr, and by separating the SFR into a background decreasing SFR measured from the minima of the SF bursts, and the bursts themselves.

\begin{figure}
    \centering
    \includegraphics[width=\linewidth]{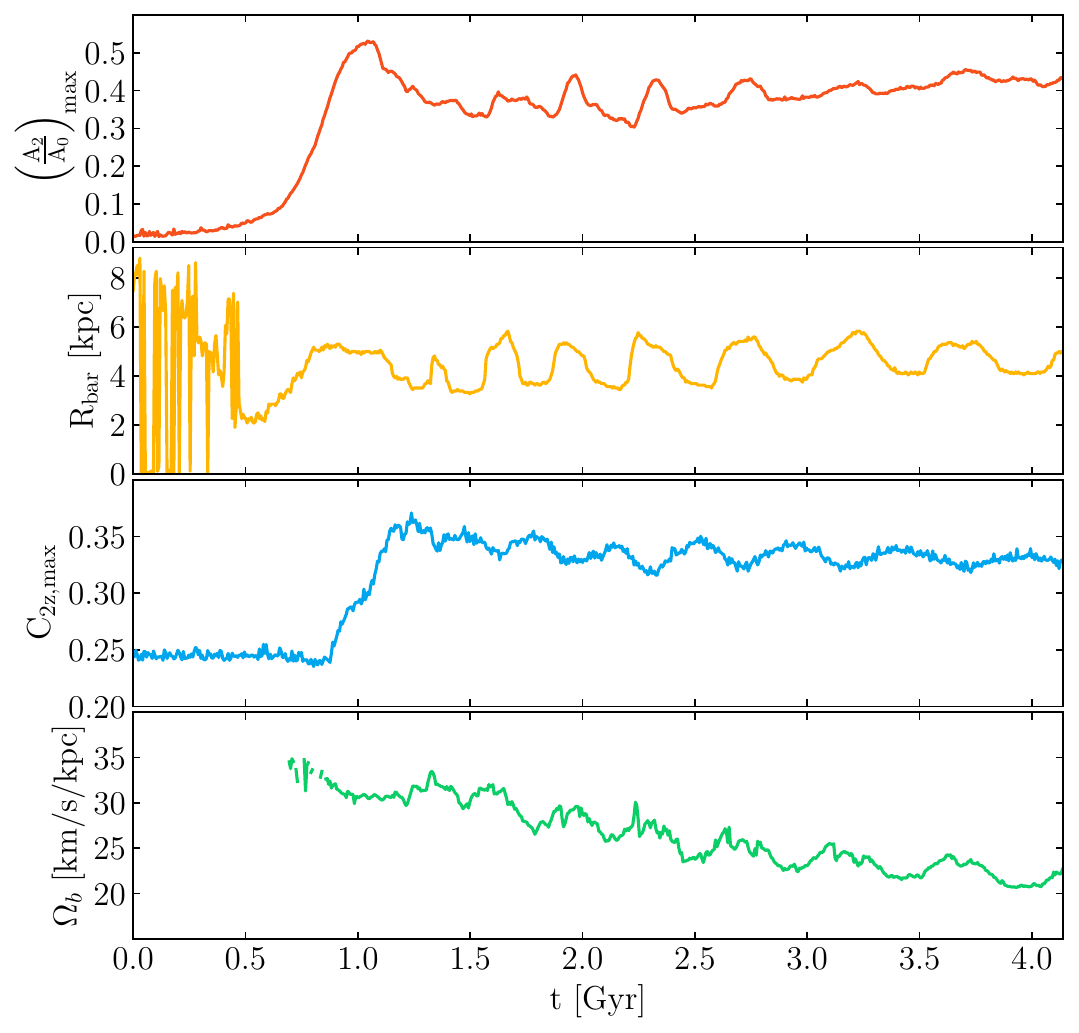}
    \caption{Time evolution of the bar and bulge properties. \textit{First panel:} Bar strength, as measured by the peak ($\frac{A_2}{A_0}$). \textit{Second panel:} Bar length (R$_{\rm bar}$), as measured by the radius at which the $\frac{A_2}{A_0}$ ratio drops to 80\% of its peak value. \textit{Third panel:} Normalised bulge strength (C$_{\rm 2z, max}$), as measured by the Fourier coefficient defined in Sect.~\ref{sec:bar_bulge_spiral}. \textit{Fourth panel:} Bar pattern speed ($\Omega_b$).
    }
    \label{fig:bar_properties}
\end{figure}

During the third phase, after bar formation, the SFR slowly decreases from $\sim$0.4~$\rm M_\odot/yr$ to $\sim$0.2~$\rm M_\odot/yr$, with fluctuation amplitudes ranging from 0.2 to 0.3~$\rm M_\odot/yr$. Six bursts can be seen in the inset plot of Fig.~\ref{fig:SFH}, each lasting a few hundred megayears. Their amplitude goes down with time, as the amount of available gas to form stars also diminishes towards the end of the simulation. These bursts are each associated with the fluctuations seen in the bar properties of Fig.~\ref{fig:bar_properties}, most clearly seen in the bar length, which extends at bar formation time (1~Gyr), B/P bulge formation time (1.25~Gyr), and in seven subsequent peaks.

\subsection{Bar-spiral arms coupling}\label{sec:bar_spiral_couple}

To understand the mechanisms driving temporal fluctuation in both bar and bulge properties (Fig.~\ref{fig:bar_properties}) and SFR (Fig.~\ref{fig:SFH}), we investigate the global asymmetry of the galaxy over time. In the upper panel of Fig.~\ref{fig:overdensity_t_r}, we show the difference $\delta\rho$ between the stellar density $\rho(|X_B|)$ along the major bar axis $\rm X_B$ (defined as stars with |$\rm Y_B$| < 0.5~kpc, with $\rm Y_B$ the minor bar axis) and the azimuthally averaged density $<\rho(R)>_\phi$, for a range of radii and over time. As expected, the simulation starts from an axisymmetric state (i.e. $\delta \rho \sim 0$ for all values of $X_B$), and an asymmetry signature appears at the time of bar formation around 1~Gyr (black dashed line). In Appendix \ref{app:overdensity_maps} we plot overdensities in face-on maps at two times to illustrate the different configurations when the bar and spiral arms are and are not connected.

Periodic signals of asymmetry after bar formation appear as elongated vertical structures (indicated with dotted vertical lines), extending the bar signature from R=5~kpc up to R=10~kpc, for short periods of time of about 100~Myr. During these events, the stellar density is enhanced along the major bar axis, leading to stronger torques exerted on the gaseous disc at further distances from the centre of the galaxy. This leads to repeated episodes of gas infall, which in turn lead to star formation bursts in the nuclear regions, albeit with a delay of a few hundred megayears, corresponding to the infall time of the gas within the nuclear regions (see the next section). These asymmetry fluctuations are caused by an overlap between the galactic bar and the spiral arms, an effect similarly reported in \cite{Debattista_2000}, \cite{Hilmi_20}, \cite{Verwilghen_24}, and \cite{Marques_25}. Multiple studies have demonstrated mode coupling between bars and spiral arms rotating at different speeds, whether theoretical~\citep{Tagger_87} or through analysis of $N$-body simulations and of power spectra~\citep{Sellwood_88,Masset_97}. As the arms rotate with a different pattern speed than the bar, they periodically overlap, leading to the observed overdensities. The spirals in our simulation are m=2 arms that extend from the end of the bar up to $\sim$10~kpc. They rotate at a pattern speed lower than that of the bar, which we measure from spectrograms, as is shown in Fig.~\ref{fig:spectrogram}.  This coupling is also seen between these reconnection events, where the density along the major bar axis outside of the bar region appears lower than the average density. At these times, spiral arms are not aligned with the bar and thus contribute to the average density, which appears higher than the density along the bar axis at high radii where the bar is weak. The last peak of fluctuations in the bar length and strength happens at t=4.14~Gyr at the very end of the simulation, and as such is not present in the SFR as a star formation burst, as the gas that is driven inwards by this last event has not yet reached the centre of the galaxy.

We note that the fluctuations in the bar pattern speed (lower panel of Fig.~\ref{fig:bar_properties}) are not an artefact of the measurement method, but are real physical fluctuations. This is made clear in \cite{Hilmi_20,Marques_25}, who measure pattern speeds in the bar region and in the spiral arms regions separately, for which they still find fluctuations. These fluctuations are anti-correlated. As the bar mode and the spiral arms mode couple, they effectively exchange part of their rotation rates back and forth. Measuring the bar pattern speed using stars within R < 3~kpc, well within the bar region, yields the same fluctuations, thus confirming their physical nature, as highlighted by the aforementioned studies.

\begin{figure}
    \centering
    \includegraphics[width=\linewidth]{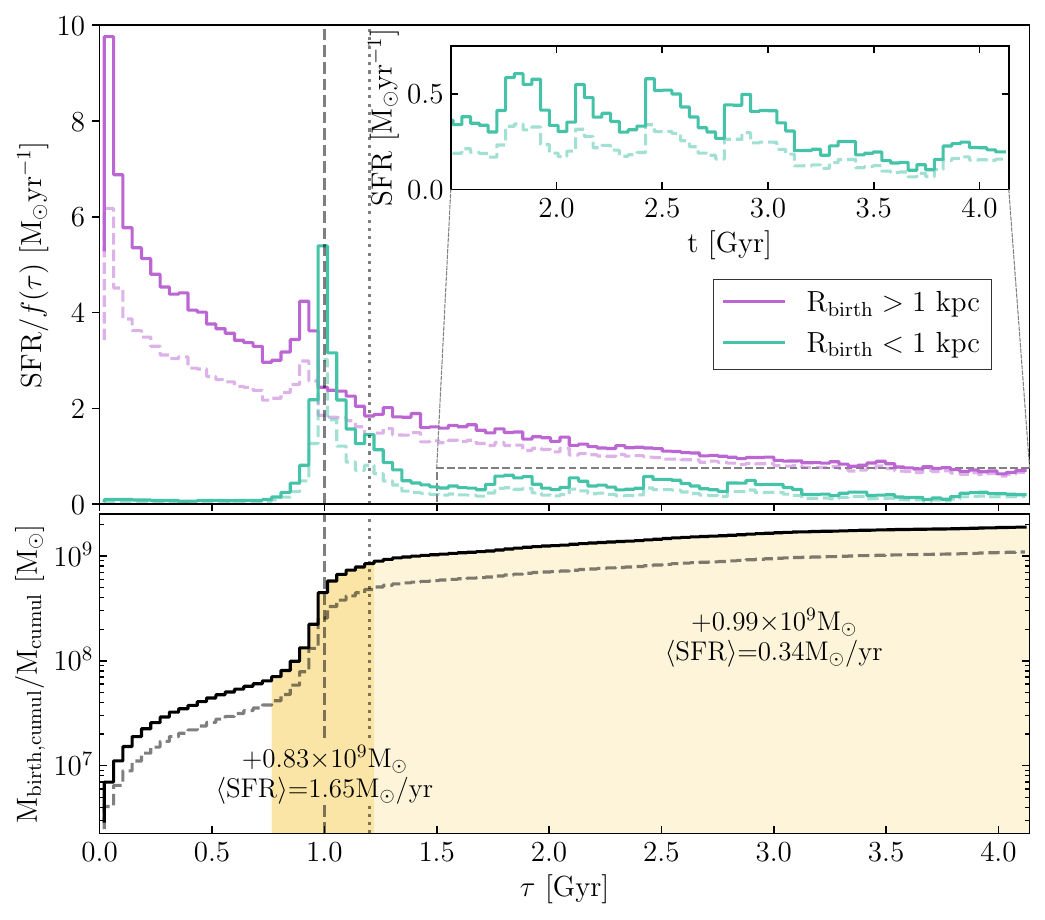}
    \caption{\textit{Top:} Star formation history for stars born at $\rm R_{birth} > 1~kpc$ (solid purple curve) and stars born within 1~kpc (solid blue curve). The dashed curves are the age histograms, weighted by the final mass of the star particles. The inset axis zooms in on the later SFH in the central regions to highlight the periodic SFR peaks. The bar formation time and bulge formation time are shown as black dashed and dotted vertical lines. \textit{Bottom:} Cumulative mass in the central 1~kpc region, weighted by initial mass in the solid curve and by final mass in the dashed curve. Mass gain during bar formation time is highlighted in yellow, and mass gain after bar formation is in light yellow. The mass gain value and mean SFR during the highlighted times are annotated.}
    \label{fig:SFH}
\end{figure}

\subsection{Nuclear structures formation}

The bar-spiral arms coupling discussed in the previous section leaves specific imprints both on the gas distribution and related star formation of the nuclear regions.
In Fig.~\ref{fig:overdensity_t_r}, we plot in the second row the distance $R(t)$ to the galaxy centre of gas particles as a function of time, and in the third row the birth radius\footnote{The birth radius of a stellar particle is defined as the distance to the galaxy centre at which this stellar particle forms.} of newly created stars as a function of their birth time.

Before bar formation, the gas particles are uniformly distributed in the inner regions (second row of  Fig.~\ref{fig:overdensity_t_r}, $t\lesssim 1~Gyr$), leading to uniformly low star formation density (bottom row of Fig.~\ref{fig:overdensity_t_r}, $t\lesssim 1~Gyr$). At bar formation time, gas from outer regions is driven to the innermost regions, and stars are formed along a range of radii, all the way to the very centre of the galaxy. The initial radius of infalling gas covers a region between R$\sim$3~kpc and R$\sim$10~kpc, corresponding to the bar end-spiral arms interface. Between 1~Gyr and 1.25~Gyr, an overdensity of gas within 100~pc forms a population of stars whose radius grows to $\sim$200~pc. Following the bar and B/P bulge formations, the gas rearranges itself into two distinct structures: the first is a nuclear ring covering radii between 300 and 400~pc, and the second is a clump covering the inner 50~pc from the galaxy centre. Both the ring and the central concentration contain high amounts of gas, which are periodically replenished by gas infall events, which are visible as vertical stripes in the second row of Fig.~\ref{fig:overdensity_t_r}, indicated with dotted vertical lines. Comparing the panels in the first and second rows, we can see that these vertical stripes appear following the bar and spiral arms reconnection, that is, after the ‘bar overdensity’ extends to distances $\rm X_B$ of up to 9-10~kpc from the centre of the galaxy. The lag between the overdensity peaks and peaks of gas infall, illustrated by the slight time differences in the dotted lines in the first and second panel, is due to the time necessary for the gas to reach the central 1~kpc.
As we checked, the gas swept up and driven inwards by the bar initially orbits at radii between 3 and 10~kpc. Contrary to the extended star formation when the bar forms, covering the entire inner 1~kpc, no stars are born within the 0.5~kpc to 1~kpc region during gas infall events at later stages. Indeed, all star formation is confined to the two nuclear structures of the ring and cluster.

In Fig.~\ref{fig:torque_plot}, we plot the time evolution of the azimuthally averaged torque radial profile of the gas, as measured by the time derivative of the z-axis component of the gas angular momentum. The regular signal seen in the bar properties aligns with regular strong negative torque values (in red), and occurs at the same time as the overdensity fluctuations displayed in the first panel of Fig.~\ref{fig:overdensity_t_r}. The region of strong torques imposed on the gas during reconnection events covers the bar/spiral arms interface, i.e. a radius slightly higher than the bar length, down to the NSD. Indeed, within the innermost regions, the gas settles into a stable ring configuration, where the gas velocity is mostly circular and  vertical torque is thus low. In Appendix~\ref{app:torque_maps} we show face-on maps of the mean torque at two different times, during and outside of a reconnection event, as well as the time evolution of mass flux through several annuli.

\begin{figure*}[!htbp]
    \centering
    \includegraphics[width=.95\linewidth]{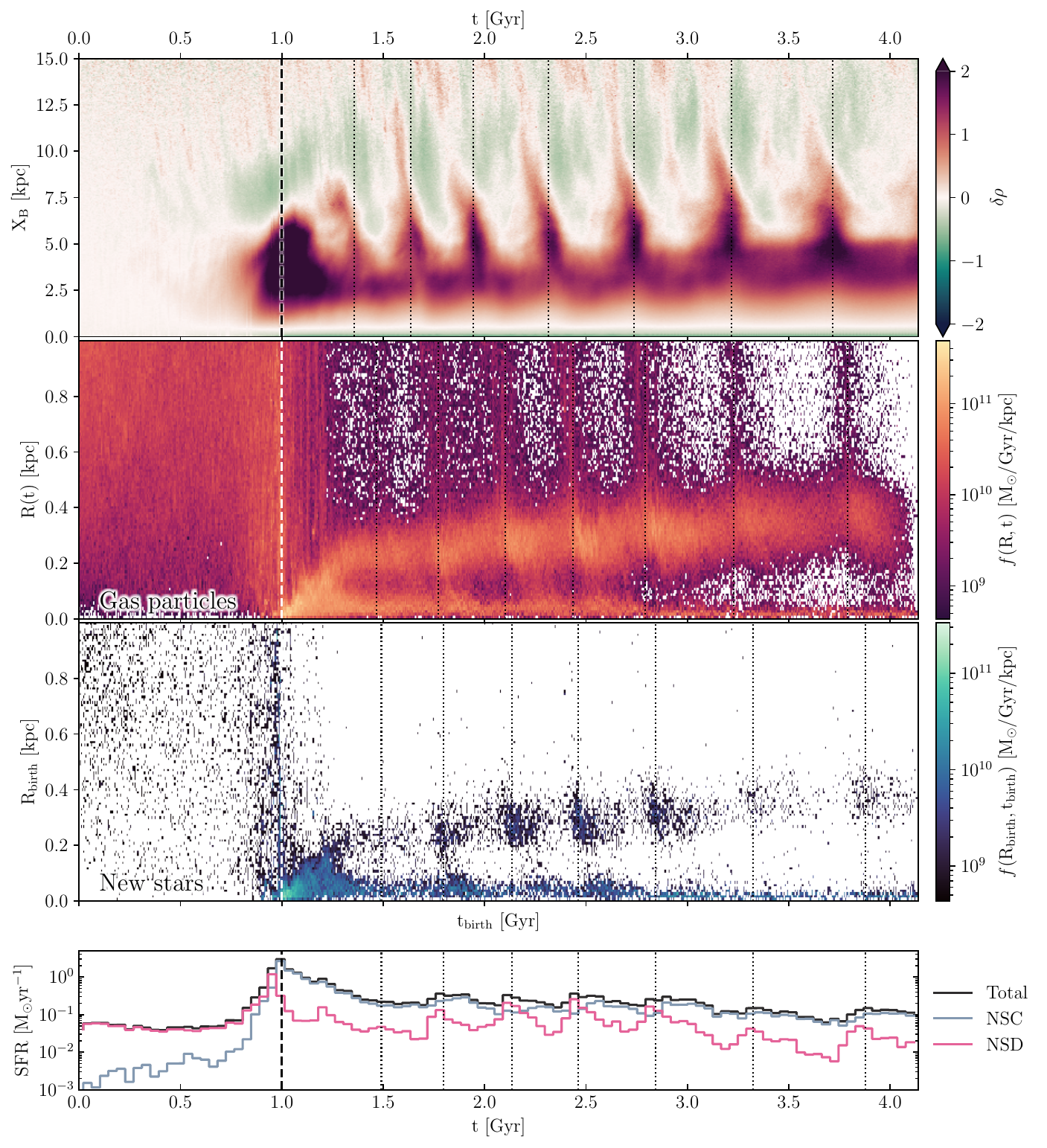}
    \caption{\textit{First row:} Overdensities $\delta \rho$ as a function of radius and time defined as ($\rho (|X_B|)$ - $<\rho (R)>_\phi$)/$<\rho (R)>_\phi$, with $\rho(|X_B|)$ as the density along the major bar axis and $<\rho(R)>_\phi$ as the azimuthally averaged density at radius R. The peaks of overdensities are indicated with dotted lines. \textit{Second row:} Orbits of gas particles (radius versus time). The peaks of gas infall are indicated with dotted lines. \textit{Third row:} Radius of birth ($\rm R_{birth}$) versus time of birth ($\rm t_{birth}$) of new star particles. The peaks of star formation in the NSD are indicated with dotted lines. \textit{Fourth row:} Star formation rates for stars born in the NSC ($\rm R_{birth} < 150~pc$) in grey, stars born in the NSD ($\rm 150~pc < R_{birth} < 1~kpc$) in magenta, and both in black. The peaks of star formation in the NSD are indicated with dotted lines. The bar formation epoch is indicated in all panels with a vertical dashed black line.
    }
    \label{fig:overdensity_t_r}
\end{figure*}

\begin{figure}
    \centering
    \includegraphics[width=\linewidth]{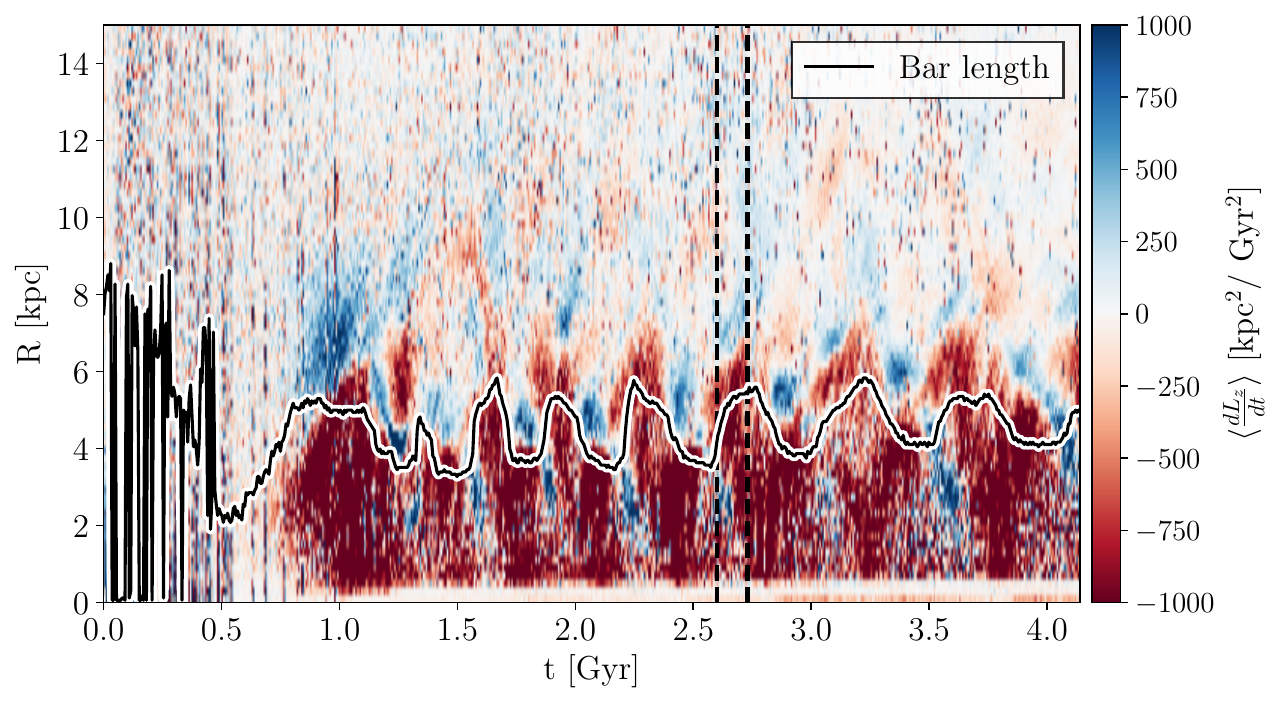}
    \caption{Mean specific torque impaired on the gas as a function of time and radius. The bar length time evolution is plotted in black to highlight the correlation between its fluctuations and high negative torque episodes in the bar region. Dashed vertical lines are times of the face-on maps in Appendix~\ref{app:torque_maps}.}
    \label{fig:torque_plot}
\end{figure}

The birth radius of newly formed stars, as a function of time, is shown in the third panel. While before bar formation, new stars form out of a uniformly distributed gas at all radii, after the bar forms, we see that new stars in the nuclear regions mostly form in two very distinct structures: in a central overdensity, the NSC, and in a region initially at about 200~pc from the galaxy centre, and with a width of about 200~pc. The peaks of star formation in the NSD region are shown in the third and fourth panels as dotted vertical lines, which are slightly delayed from the peaks of gas infall. The second overdensity of young stars corresponds to the nuclear ring, also visible in Fig.~\ref{fig:face_on_maps}, and traces the outer edge of the nuclear disc. The nuclear disc itself is not apparent in this panel because only stars formed within the time interval between $t$ and $t+\Delta t$ are shown. As a result, the region between the nuclear ring and the NSC appears nearly empty. However, if all stars formed in the nuclear region since the beginning of the simulation were included, this intermediate region would be populated, revealing the full extent of the nuclear disc. As reported in \cite{Nogueras_lara_23b}, the MW NSD displays an age gradient, with younger stellar populations in its outer regions than in the inner ones. Our findings are consistent with these observations, as star formation primarily occurs in the ring at the outskirts of the NSD, while the inner region is populated by stars having migrated there some time after their birth. Our results are similar to those of \cite{Sormani_24}, which find that the star formation in the nuclear regions occurs in rings of growing radii.

From the third row of Fig.~\ref{fig:overdensity_t_r}, we can also see that the nuclear ring and the NSC gradually diverge with time, with the nuclear ring growing in size, from a mean radius of 200~pc at its formation at t=1.25~Gyr, to 400~pc at t=4.14~Gyr. The NSC, in turn, follows the opposite trend becoming increasingly compact, although its evolution below the gravitational softening scale cannot be reliably resolved.

As shown in the bottom panel of Fig.~\ref{fig:bar_properties}, the bar slows down over time, leading to higher radii for the corotation ($\Omega=\Omega_b$), inner Lindblad resonance (ILR; $\Omega=\Omega_b+\frac{\kappa}{2}$) and outer Lindblad resonance ($\Omega=\Omega_b-\frac{\kappa}{2}$). As measured from the potential under the epicyclic approximation, the radius of the ILR grows from 1~kpc at t=1~Gyr to 2.6~kpc at t=4.14~Gyr. In the literature, rings are often associated with the presence of an ILR (see, for example, \cite{Combes_88} or \cite{Combes_01} for a review on gas flows in barred galaxies). In our simulation, we find that – regardless of how we estimate the ILR (via the epicyclic approximation or via frequency analysis) – the ring always appears within the ILR of the primary bar. This would seem to support the analysis presented in \cite{Sormani_24}, which proposes that rings are an accumulation of gas at the inner edge of a wide gap that forms around the ILR. In our simulation, we do see the formation of such a gap inside the ILR, particularly evident in the gas distribution. While the behaviour of $\Omega$ in the inner Galaxy determines the position of the ILR, this position is well defined only for nearly circular motion in a razor-thin disc. In the case of non-planar orbits, or more generally in a three-dimensional system, the ILR location derived using the standard approach may not correspond to its true location, especially when the vertical excursions of stars are comparable to the disc thickness. This picture is further complicated by the presence of the NSD component itself. Further analysis, which goes beyond the scope of this article, would nevertheless be necessary to understand the location of this ring, its evolution over time, and its connection – if any – with the nuclear bar resonances.

In the bottom panel of Fig.~\ref{fig:overdensity_t_r}, we plot the SFRs computed for stars born within $\rm R_{birth} < 150~pc$ (grey curve, the NSC-dominated region) and within $\rm 150~pc < R_{birth} < 1~kpc$ (magenta curve, the NSD-dominated region). While they both display regular star formation bursts, which are associated with the bar-spiral arms coupling, as seen by comparing with the top panel, the NSC SFR is offset, with its SFR peaks occurring a few hundred megayears after those of the NSD SFR. This is also true for the main SFR peak at bar formation, which occurs at t=0.95~Gyr for the NSD. The total star formation in the nuclear regions is dominated by the SFR in the NSC, which is $\sim$5 times higher than the SFR in the NSD after bar formation. While both the NSC and NSD experience periodic bursts of SFR, these bursts appear stronger (relative to the background or average SF) in the NSD than in the NSC. Indeed, the NSC has a background SFR of $\sim0.15\rm~M_\odot/yr$ at t=1~Gyr, reaching $\sim0.05\rm~M_\odot/yr$ at t=4.14~Gyr. The bursts increase this background SFR by 0.10 to 0.15$\rm~M_\odot/yr$, i.e. by a factor of 2 on average. For the NSD, the background SFR is $\sim0.05\rm~M_\odot/yr$ at t=1~Gyr and $\sim0.005~\rm M_\odot/yr$ at t=4.14~Gyr. The bursts increase this background SFR by 0.15 to 0.20$\rm~M_\odot/yr$, i.e. by a factor of 5 on average. This is consistent with the third row of Fig.~\ref{fig:overdensity_t_r}, which shows new stars in the NSD only during gas infall events, while the NSC continuously forms stars, at a higher rate during these events.

In Fig.~\ref{fig:SK_law}, we show the SFR surface density, defined as the SFR per unit area, as a function of the gas surface density for both the NSD and the NSC at different times. The NSC has a higher star formation efficiency at any time, with a depletion time of $\rm\tau_{depl}\sim10^8~yr$, compared to a depletion time of $\rm10^9~yr$ for the NSD. In both cases, the depletion time is measured as the ratio of total gas surface density to the stellar formation rate density; thus, it is a depletion time for all gas. We explain this by the fact that the NSC retains gas in between gas infall events and is thus able to form stars sustainably, while the NSD quickly depletes its gas content, either by forming stars shortly after gas deposition or by transferring some of its gaseous material to the NSC. During gas inflow events, the depletion time, $\tau_{\rm depl}$, decreases in the NSD. In the NSC, instead, $\tau_{\rm depl}$ follows a smoother, quasi-monotonic decline. As we previously showed, out of the two nuclear structures, the NSD displays a star formation evolution that is more influenced by gas infall events than the NSC, which has a more continuous SFR, as the former quickly depletes its gas content while the latter does not have time to convert its entire gas reservoir in between gas infall events. Hence, the depletion time will vary more greatly in the NSD, with added gas driving a decrease in the depletion time, followed by an increase when this gas has been turned into stars.

\begin{figure}
    \centering
    \includegraphics[width=\linewidth]{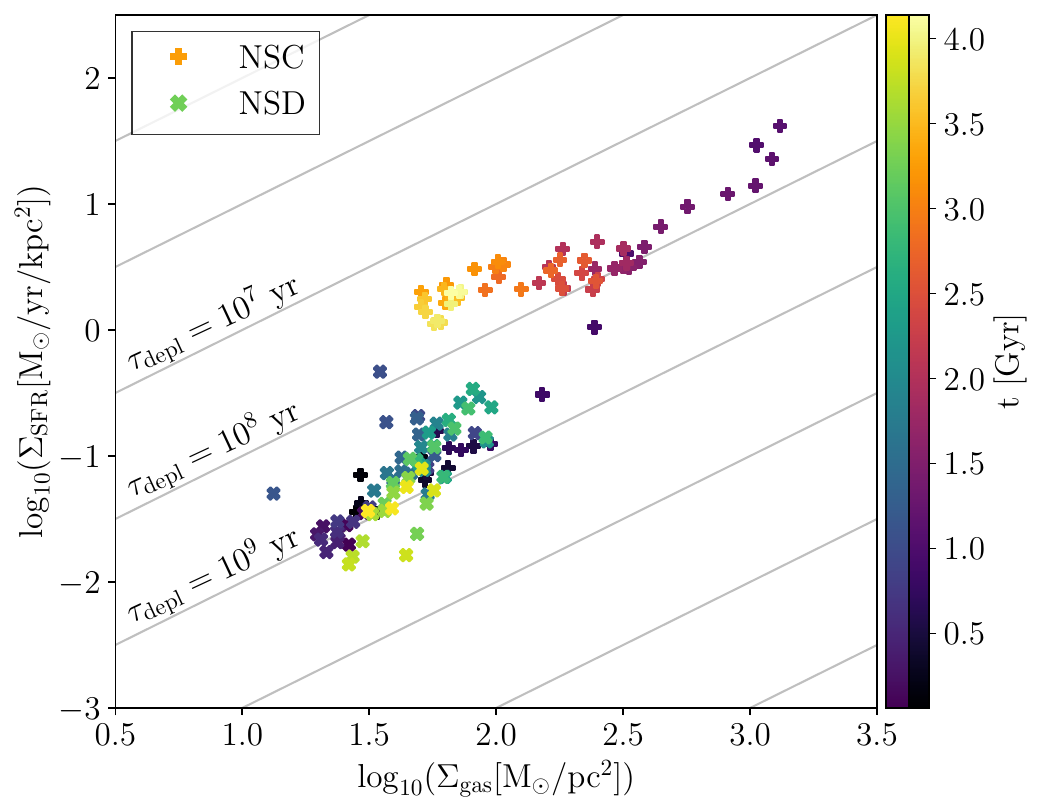}
    \caption{Gas surface density versus SFR surface density for the NSC ("$+$" symbols) and NSD ("$\times$" symbols). Colours become lighter as time evolves. Lines of similar depletion time are shown in light grey.}
    \label{fig:SK_law}
\end{figure}

\section{Discussion}\label{sec:discu}

\subsection{Bar-spiral arms interaction}

The interaction of galactic bars and spiral arms has been studied in previous works. \cite{Marques_25} analysed three models of Milky Way-like galaxies, including one isolated simulation and two cosmological ones, and found that variations in the bar length drive periodic bursts of star formation at the bar ends. During these episodes, the SFR increases by a factor of $\sim$ 3–4 above the baseline value. In the inner regions, our results show an SFR enhancement by a factor of more than two following reconnection events. Using the same models in a cosmological context, \cite{Hilmi_20} found variations of bar lengths of 1.5~kpc in the highest case, overestimating the true bar length by up to 100\%. Our bar extends to larger radii than theirs; as such, the relative amplitudes are smaller; however, the absolute amplitudes are of the same order of magnitude, reaching 2~kpc. In addition, they find variations in the bar strength, correlated with variations in bar length. We also find that the bar pattern speed fluctuates following bar-spiral arms interactions; however, \cite{Hilmi_20} found that their pattern speeds were mostly anti-correlated with bar length and strength, whereas we find they are closely correlated. Their fluctuations are correlated with SFH fluctuations, independently of the simulation analysed. We note that overall, we find the expected anti-correlation between bar strength and bar pattern speed, as the bar in our simulation grows stronger with time and decelerates. However, the small timescale fluctuations in bar strength and pattern speed, which are caused by bar-spiral arms coupling and not secular evolution of the bar itself, are correlated in our case. When measuring the corotation radius of the bar, we find the expected anti-correlation with the bar pattern speed fluctuations. Similarly, \cite{Verwilghen_24} find fluctuations in the $\rm A_2$ of their G053 model (their Fig.~7 and 8), which they associate with bar-spiral arm interactions and that impact the SFR of the central 1~kpc.
Multiple studies of the  Milky Way non-axisymmetries \citep{Bissantz_03,Dias_05,Gerhard_11} found  
spiral arms pattern speed estimates significantly lower than that of the Galactic bar. If the bar and spiral arms do not share the same pattern speed, their relative angle necessarily varies in time. Indeed, in a recent work using \textit{Gaia} DR3 data, \cite{Khalil_25} measured a m=2 spiral arm pattern speed of 13.1~km/s/kpc, and concluded that the bar recently disconnected from it.

In their study of a simulation of an isolated galaxy, \cite{Fanali_15} find a burst of gas inflow at bar formation time, followed by very little gas inflow after that. Bar-spiral arms interaction effects thus seem low to non-existent, contrary to our results. A major difference between the two simulations is the absence of star formation in \cite{Fanali_15}'s model. As such, we expect spiral arms to be weaker in their models than ours, being primarily traced by young stars. Their interaction with the bar would hence have a much weaker effect on the dynamics of the gaseous component.

Similarly, \cite{Pettitt_20} studied a suite of simulated isolated galaxies and the evolution of the ISM and star formation in the mid to outer regions of the galaxy along the spiral arms, under different underlying fixed potentials. Their potentials have axisymmetric components, to which they add a bar potential and/or a spiral potential. The different pattern speeds given to their bar and spiral arms lead to regular overlaps at 'beat' frequencies, at which they find SFR bursts (their fig. 6), similar to our findings. As in our case, they find that these bursts are regularly spaced and of amplitudes of a few $0.1~\rm M_\odot/yr$, decreasing with time as the gas reservoir empties. In their study, \cite{Seo_14} also find that spiral arms can be efficient at transporting gas to the central regions and enhance star formation in nuclear rings episodically, if they are rotating at a slower rate than the bar.

Once the gas loses angular momentum and has reached the nuclear ring, an additional mechanism is needed for it to reach the central-most regions. \cite{Shlosman_89} invoke the presence of inner bars as a means to drive secondary gas infall towards the very centre of galaxies. In our simulation, new stars created in the nuclear ring rearrange into a bar-like structure on top of the NSD (see the middle lower panel of Fig.~\ref{fig:face_on_maps}), which can act as the final link between the ring and the NSC. \cite{Emsellem_15} and, more recently, \cite{Kobayashi_26} both describe another mechanism by which stellar feedback is able to uplift gas filaments at high galactic heights, in a fountain-like manner, permitting it to reach smaller radii. In our simulation, we do not find evidence of gas gaining height at the nuclear ring radius.

By studying 57 disc galaxies from the EDGE-CALIFA survey, \cite{Yu_22} find a correlation between a central concentration of molecular gas and both the presence and strength of bars, as has been found in other studies (see e.g. \citealt{Kuno_07}). Additionally, they find a correlation with the strength of non-axisymmetric components in general, i.e. bars and spiral arms. In these galaxies, star formation closely follows molecular gas concentration, thus highlighting a link between bars and spiral arms and an enhanced star formation in the central regions.

\subsection{Star formation in the nuclear regions}

The scale of the nuclear ring is in agreement with results from the PHANGS-ALMA survey \citep{GLEIS_26}, finding a median nuclear ring radius of 400~pc. In our simulation, the nuclear ring has a final mass of $\rm\log_{10}(M/M_\odot)=7.43$, which is an order of magnitude lower than the PHANGS-ALMA value of $\rm\log_{10}(M/M_\odot)=8.1$. As mentioned earlier, our simulation is integrated in isolation, with no gas accretion events; thus, it represents a lower bound on gas masses. 

The nuclear regions host a variety of structures represented by different stellar components, forming at different times. When using the previously defined delimitation of R=150~pc between the two star forming regions in the inner nuclear region, we obtain a mass for the NSD of $\rm 3.6\cdot10^{8}~M_\odot$ and of $\rm 6.8\cdot10^{8}~M_\odot$ for the NSC, in disagreement with the reported values of a few $\rm 10^9~M_\odot$ for the MW NSD \citep{Sormani_22} and a few $\rm 10^{7}~M_\odot$ for the MW NSC \citep{2014A&A...566A..47S, 2016ApJ...821...44F}.
However, the combined mass of the NSC and NSD within 400~pc is of $\rm\sim10^9~M_\odot$, close to the combined reported value of the MW NSD and NSC. While its formation scenario is well understood in our simulation, the mass and internal structure of the NSC cannot be resolved, on account of the resolution set by the softening length being 50~pc and the smoothing length being 100~pc. Thus, while the total matter content is in agreement with MW observations, its distribution within the central 100~pc as well as the internal structure therein cannot be properly reproduced at this resolution.

Finally, we also calculated the mass of these structures by including pre-existing old stars, to estimate possible bulge contaminations. For a boundary radius of 10~pc, and a maximal height of $|z|=\rm 10~pc$, we find a mass of the NSD of $\rm 2.5\cdot10^9~M_\odot$ and a mass of the NSC of $\rm 3.4\cdot10^7~M_\odot$, i.e. an overestimation of 143\% for the NSD and of only 2\% for the NSC. This overestimation scales with the maximal height used, as the measurement captures more and more interlopers from the bulge.

As shown by \cite{Baba_20}, the main peak of SFR in the NSD region is a tracer for the bar formation epoch, as a significant amount of gas is funnelled to the central regions following the event \citep{Roberts_79,Combes_1985,Shlosman_89,Athanassoula_92,Combes_93,Seo_19}. This fact has first been used in \cite{Gadotti_15} to date the bar of NGC 4371, followed by \cite{De_Sa_Freitas_23}, who developed a methodology using the SFH of the NSD provided by the TIMER survey \citep{Gadotti_19} to date bars, and applied it to date the bar of NGC 1433. In a similar fashion, \cite{Sanders_24} dated the time of formation of the Milky Way bar, finding an age of the bar $\sim 8~$Gyr. This technique has been further used in \cite{de_sa_Freitas_25} to date bar formation times for 20 nearby galaxies.

During the quiescent phases by the end of the simulation, the SFR matches the SFR reported from the PHANGS-ALMA survey \citep{GLEIS_26}, finding a median SFR value of 0.21$~\rm M_\odot/yr$ across their sample of galaxies. As previously mentioned, our simulation does not include additional gas accretion from filaments, and as such features a decrease in the background SFR from 6 $\rm M_\odot/yr$ at the start of the simulation to 1 $\rm M_\odot/yr$ at the end. Further work will consider gas accretion, which should help explain the recent star formation bursts observed.

\cite{Nogueras_Lara_20}, using the GALACTICNUCLEUS survey \citep{GalacticNucleus}, found a SFH in the central regions of the Milky Way with a significant portion of stars with ages $>8~\rm Gyr$, followed by a quiescent phase, with a notable peak $\sim1~\rm Gyr$ ago. They conclude that bar-induced inflows have been mostly inefficient at rejuvenating the inner regions following the initial inflow. They also measure a recent increase of SFR in the last 10 to 100~Myr, as has \cite{Schodel_20}. In subsequent studies broadening the scope to more outer regions of the Milky Way NSD \citep{Nogueras_Lara_22a,Nogueras_lara_24b}, as well as in \cite{Schodel_20}, they find that these outer regions show a significant portion of intermediate-aged stars. They conclude that these observations are consistent with an 'inside-out' formation scenario for the NSD. \cite{Sanders_24}, using Mira variables and a Period-Age relation, built an SFH for the Milky Way NSD and associated its main peak with the bar formation epoch. Their SFH shows variations at later times, which are not seen in the bar-bulge/disc SFH, as is the case in our models. For a complete review of the Milky Way NSD SFH, see \cite{Schultheis_25}.

As for the Milky Way NSC, observations show that, similarly to the NSD, most stars are older than 5~Gyr, with a recent increase of number of young stars \citep{Blum_03,Pfuhl_11}.
As mentioned in the review of NSCs by \cite{Neumayer_20}, these observations of very young stars in the Milky Way NSC go against the scenario of formation through star cluster infalls \citep{Neumayer_20}, as the latter would necessarily lead to an upper limit of ages dictated by the characteristic time needed for star clusters to form and subsequently fall within the central regions. Observations show that the Milky Way NSC contains metal-rich stars, in higher proportion than in the Milky Way inner bulge \citep{Schultheis_19,Feldmeier_krause_20}, a metallicity gradient of $\sim-0.1\pm0.02$~dex/pc \citep{Schultheis_26}, as well as abundances distinct from globular cluster-like signatures 
(\cite{Ryde_25} for the MW NSD and \cite{Nandakumar_25} for the MW NSC), which also provides evidence against the cluster-infall formation scenario. In our simulation, the NSC is star-forming throughout its entire evolution, fed by regular gas inflows. Thus, it appears much younger than the outer regions, as well as younger than the NSD, which has a lower SFR. Surveys of nearby late-type galaxies NSCs reveal multiple distinct stellar populations, supporting an episodic and extended star formation \citep{Walcher_06,Rossa_06,Carson_15}.

Just as \cite{Gadotti_15}, \cite{De_Sa_Freitas_23} and \cite{Sanders_24} used observed SFH to derive bar formation epochs, our findings suggest a possible use of the nuclear regions SFH to constrain the evolution of spiral arms and their interaction with the galactic bar. Indeed, the period of these SFR bursts corresponds to the difference in pattern speeds between the bar and spiral arms (see Appendix~\ref{app:spectrogram} for a derivation of the spiral arms pattern speed in our simulation). Coupled with existing constraints on the bar pattern speed history and upcoming surveys like MOONS \citep{Cirasuolo_20}, this could potentially put constraints on the spiral arms history, which has seldom been addressed for lack of dynamical signatures. Recently, \cite{Spitoni_23} and \cite{Barbillon_25} highlighted the effect of spiral arms evolution on the radial and azimuthal variations of different chemical abundances, whose producers do not have the same lifetime~\citep[see also][]{2018A&A...611L...2K, 2023A&A...671A..56K}.

\cite{Kwak_26a} studied the conditions for the formation of nuclear disks and rings with a varying central bulge mass through self-consistent hydrodynamics simulations with multiphase ISM. They find that a central mass concentration is necessary for the build-up of NSDs, which is otherwise disrupted by stellar feedback, contrary to our findings, which do not include such a component. The SFH of their central regions also fluctuates, due in part to bar-spiral arm interactions but mostly due to local feedback events. The properties of the NSDs and NSCs they form match ours in terms of ratio kinematics, with a ratio $v/\sigma$ of $<0.7$ for NSCs and 1.2-1.7 for rings, and a range of disk sizes between 0.25 and 0.75~kpc. In their study, \cite{Seo_19} found that nuclear disks could still form without a central spheroidal bulge in their simulations, using weaker feedback than in \cite{Kwak_26a}. In \cite{Kwak_26b}, the authors run higher-resolution simulations of the SMUGGLE ISM model, in which they also find that feedback regulates the formation and evolution of the NSC and NSD, and, as they previously found, that the resulting size of their NSD hints at a bulge-to-disc mass ratio lower than the 0.045 value they adopt. 

Finally, \cite{Spitoni_26} built a chemical evolution model of the formation of the Milky Way NSD, in which they find that a single bar-induced gas inflow cannot explain the presence of metal-poor stars in the metallicity distribution function of the NSD. They concluded that either the gas that formed the NSD must have been diluted from gas in the thin disc or from accretion events in order to populate the low metallicity tail of the MDF or the NSD metallicity distribution functions are contaminated by bulge stars. In our simulation, following the initial gas inflow event caused by the bar formation, subsequent bar-spiral arm interactions drive gas from radii up to 10~kpc to fall inwards. Coupled with an initial radial metallicity gradient, this provides another viable way to dilute the gas that forms the NSD, since material from the  outer disc would have lower metallicity than the inner-disc material used at bar formation to form the NSD.

This breadth of numerical studies and formation paths of nuclear structures hints at the importance of the physical recipes used to simulate these regions. A different sub-grid model, resolution, initial conditions and inclusion in a cosmological context could lead to stronger or weaker gas inflows, mass build-up and size of the central structures, as the main mechanism in our simulation depends on the properties of the bar and spiral arms, which might vary under different contexts and physics. Nonetheless, our results, consistent with the recent works by \cite{Hilmi_20} and \cite{Marques_25}, highlight a commonplace occurrence for bar-spiral coupling.

Including a cosmological context that may provide gas accretion, replenishing the galaxy's gas reservoir, might also affect the observed SFR bursts in the central region. These bursts could be sustained for a longer time as the gas in the outer region would deplete more slowly than in an isolated galaxy context, and could also be of higher amplitude. On the other hand, if the gas accretion onto the galaxy is bursty, the periodic SFR bursts of the NSC/NSD might get disturbed, as  bar-spiral arm reconnections might not be the main drivers of gas inflows.

\section{Conclusion}\label{sec:conclu}

We have presented an \textit{N}-body plus hydrodynamics simulation of an isolated Milky Way-like galaxy evolved from an initial pre-existing stellar disc and a gas disc. The discs quickly form a galactic bar that buckles to form a B/P bulge. The formation of these non-axisymmetric features leads to gas being funnelled to the central sub-kiloparsec regions, where star formation occurs. The inner regions quickly rearranges into an NSD, a nuclear bar, and an NSC. Their SFH displays a major burst at t=1~Gyr with an SFR of 5~$\rm M_\odot/yr$, compared to a previous rate close to zero. Subsequently, the SFH in the inner regions does not decrease monotonically as the disc SFH does, but instead it shows fluctuations lasting a few hundred megayears and separated by several hundred megayears. The amplitude of the star formation bursts is of 0.2 to 0.3~$\rm M_\odot/yr$, effectively doubling the SFR during these short periods of time.

On the large scale, we measured the bar length, strength, and pattern speed as well as the bulge strength. We found that these properties all display the same SFR fluctuations that were seen in the nuclear region. We showed how these fluctuations are linked to the interaction between the galactic bar and the spiral arms. As the spiral arms rotate at a different pattern speed than the bar, these two structures regularly overlap, leading to a bar that appears longer, stronger, and faster. During the reconnection events, the increase in stellar density along the bar, reaching higher radii, imparts torques on the gas disc, triggering secondary gas infall events and leading to star formation bursts in the NSD and NSC, with a stronger effect on the SFH of the NSD. 

Our simulation suggests that as a consequence of these regular overlaps, nuclear regions of barred galaxies can rejuvenate. This finding may have implications for their build-up over time as well as for the properties of their stellar populations.

Finally, the results presented in this article relate to a single simulation. In the future, it will be important to carry out a statistical analysis of a set of hydro--N-body simulations in order to understand the frequency with which nuclear structures form in barred galaxies, their evolution over time, their chemical properties, and their relationship with the stellar populations of the bulge and the disc. Notably, a different initial gas fraction than that used in this work -- 20\% -- would provide valuable information on the conditions under which the presence and strength of a bar and spiral arms can lead to strong enough gas inflows to build up nuclear structures.

\begin{acknowledgements}
We thank Ivan Minchev and Eric Emsellem for their helpful comments and discussion. T.~Boin and PDM thank the Ecole Doctorale Astronomie et Astrophysique d’Ile de France for funding this thesis project. S.K. acknowledges support from the Deutsche Forschungsgemeinschaft under the grant KH~500/2-1.\\
The research in this paper made use of the SWIFT open-source simulation code (\url{http://www.swiftsim.com}, \cite{schaller_18}) version 2025.04.\\
This work has made use of the computational
resources allocated on Irene Rome by GENCI, through the DARI project A0180410154.\\
The following python libraries were used for this study:  \texttt{numpy} \citep{numpy} and \texttt{matplotlib} \citep{matplotlib}.
\end{acknowledgements}

\bibliographystyle{aa_url}
\bibliography{bibliography.bib}

\begin{appendix}
\nolinenumbers

\section{Initial rotation curve}\label{app:rotation_curve}

\begin{figure}[!htbp]
    \centering
    \includegraphics[width=\linewidth]{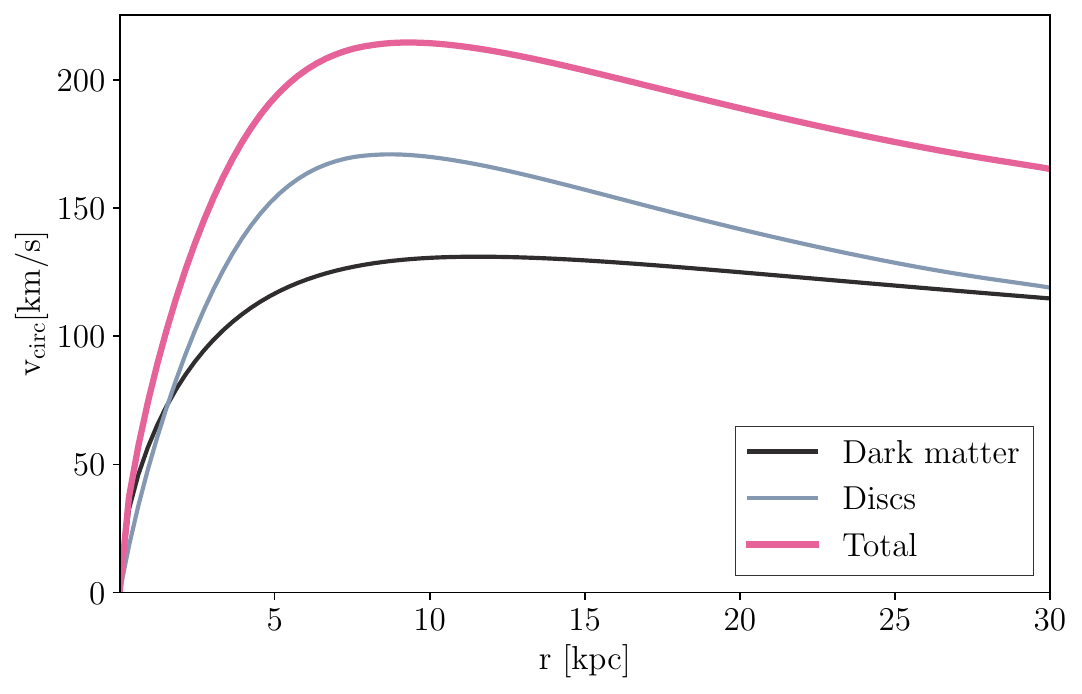}
    \caption{Rotation curve of the initial setup of the simulation, from the discs potential (blue curve), dark matter halo potential (black curve) and total potential (pink curve).}
    \label{fig:rotation_curve}
\end{figure}

\section{Additional bar length estimations}\label{app:bar_length}
In this section we measure the bar length time evolution using two additional methods in addition to the $\left(\frac{\rm A_2}{\rm A_0}\right)_{\rm max}$ criterion shown in Fig.~\ref{fig:bar_properties}. The first one is described in \cite{Aguerri_2000} and based on the contrasts between the bar and interbar azimuthal luminosity profiles. Based on their definition, we thus define the bar length as the maximal extent for which the following condition is met:

\begin{equation}
    \left(I_{\rm b}/I_{\rm ib}\right) > \frac{\left(I_{\rm b}/I_{\rm ib}\right)_{\rm max}-\left(I_{\rm b}/I_{\rm ib}\right)_{\rm min}}{2} + \left(I_{\rm b}/I_{\rm ib}\right)_{\rm min},
\end{equation}

\noindent where $I_{\rm b}=I_0+I_2+I_4+I_6$, $I_{\rm ib}=I_0-I_2+I_4-I_6$ and $I_{\rm m}$ is defined in \cite{Aguerri_2000}\footnote{A square root was added to the definition of $I_{\rm m}(r) = \sqrt{A_{\rm m}^2(r) + B_{\rm m}^2(r)}$ and that was missing in the original publication, in order to preserve dimensional homogeneity with the definition of $I_0(r)$.}. To build luminosity profiles from our simulation snapshots, we assume a mass-to-light ratio of $2\Upsilon_\odot$. The bar length, as measured through this method, is shown in Fig.~\ref{fig:bar_length} in green. Once the bar forms, this estimation shows similar fluctuations to the ones observed in the $\left(\rm A_2/\rm A_0\right)_{\rm max}$ method (in purple).

Finally, we also estimated the bar length using the radial profile of the phase-angle $\phi_2$ of the m=2 Fourier mode. Correspondingly, the bar length is the region over which the profile does not vary over 3°, plotted in Fig.~\ref{fig:bar_length} in blue. While the fluctuations seen in the first two methods are also seen in this estimate, the latter displays peaks at the minima shown in the other two. These occur in between two reconnection events, when the bar is perpendicular to the spiral arms, and can be explained as follows. Following a reconnection event, the bar ends overtake the spiral arms they were connected to, producing a break in the phase-angle profile in a trailing arm fashion. Shortly before the following reconnection event, the phase-angle profile is broken in a leading arm fashion as the bar ends approach the spiral arms ahead. In between those configurations, when the bar ends are perpendicular to the spiral arms, the phase-angle profile transitions from displaying a trailing-break to a leading-break, leading to a short period of no break, seen in the bar length signal as a short-lived increase in value.

Our results are in agreement with \cite{Ghosh_24}, namely that the $\rm\left(A_2/A_0\right)$ profile criterion (their $\rm R_{bar,3}$) tends to overestimate the bar length in the presence of spiral arms, and that the method of \cite{Aguerri_2000} (their $\rm R_{bar,5}$) yields similar results. In addition, we find that the $\Delta\phi_2(R)<3°$ criterion (their $\rm R_{bar,1}$), in the presence of spiral arms, yields more stronger and faster bar length fluctuations than the other ones.

\begin{figure}[!htbp]
    \centering
    \includegraphics[width=\linewidth]{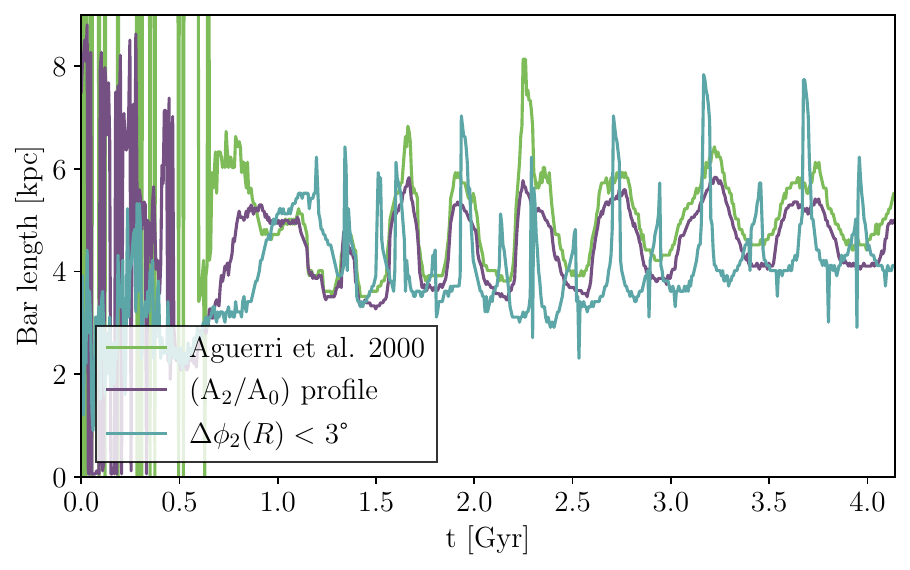}
    \caption{Bar length as measured by different methods as a function of time. Bar formation occurs at t$\sim1~$Gyr.}
    \label{fig:bar_length}
\end{figure}

Notwithstanding these secondary peaks seen in the phase-angle bar length signal, all methods display bar length fluctuations of similar amplitudes and period. Hence, these fluctuations are not a by-product of a specific bar length measurement method employed, but reflect a change in the bar properties, that is materialised in the regular gas infall events observed. Whether these fluctuations reflect a dynamical change in the stars composing the bar or they simply are a by-product shared by all bar length measurement methods when spiral arms overlap with the bar, the resulting torques impaired on the gas and subsequent mass build-up in the central regions are indeed caused by this overlap.

\section{Overdensity face-on maps}\label{app:overdensity_maps}

\begin{figure}[!htbp]
    \centering
    \includegraphics[width=.975\linewidth]{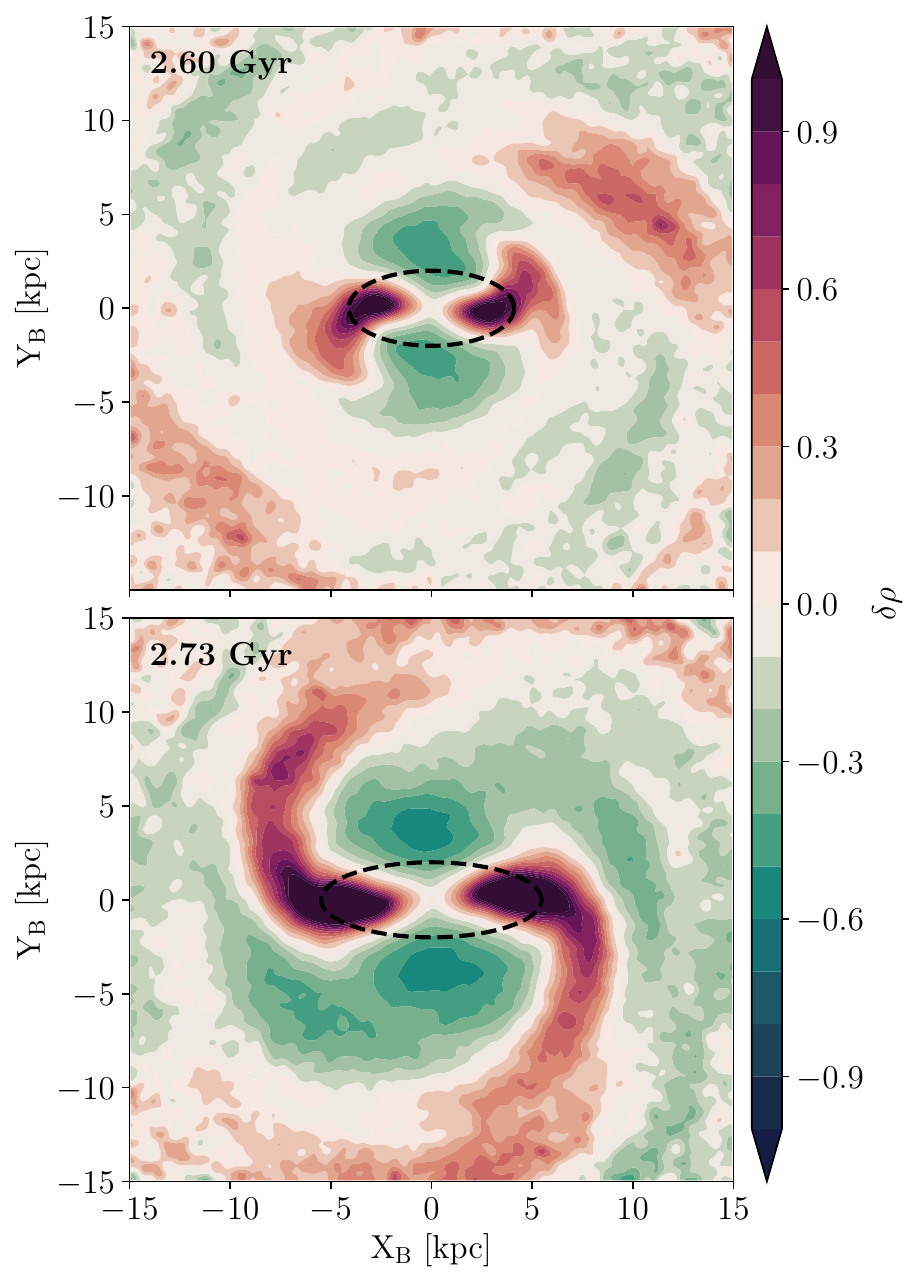}
    \caption{Overdensity face-on maps $\delta\rho$, for two different times. The maps are oriented along the major and minor bar axes $\rm X_B$ and $\rm Y_B$. The bar region is shown in dashed ellipses, where the width along the $\rm X_B$ axis is the bar length at that time.}
    \label{fig:overdensity_maps}
\end{figure}

To illustrate the configurations when the bar and spiral arms reconnect and are out of phase, we plot in Fig.~\ref{fig:overdensity_maps} the overdensities, as defined in a similar fashion as in Sect.~\ref{sec:bar_spiral_couple}, but for any position in the galactic plane, $\rm\delta\rho=\frac{\rho(X,Y)-\langle\rho(R)\rangle_\phi}{\langle\rho(R)\rangle_\phi}$. In the top panel, the spiral arms are not connected with the bar, and as such the overdensity caused by the bar extends to R$\sim$5~kpc. In the bottom panel, spiral arms are connected to the bar, leading to an extended signature of overdensity, extending to R$\sim$8~kpc. This difference can also be seen in the top panel of Fig.~\ref{fig:overdensity_t_r} at those times.

\section{Torque maps and mass influx}\label{app:torque_maps}

In Fig.~\ref{fig:torque_maps}, we plot face-on maps of the mean specific torque impaired on the gas, for two different times, at t=2.6~Gyr in between two reconnection event, and at t=2.73~Gyr when a bar-spiral arms reconnection occurs. In the latter case, the torques in the bar region are mostly negative, with gas infall occurring following the reconnection event, along the whole bar region and up to the bar-spiral arms interface at 6~kpc (see Fig.~\ref{fig:torque_plot}). In the former case, the infalling gas caused by the previous reconnection event has already reached the central region, and thus the bar region is mostly populated with the remaining gas, which has either positive or zero vertical torque (the top panel of Fig.~\ref{fig:torque_maps} displays a fainter red signal of negative torques as the infalling gas has been mostly depleted by then).

\begin{figure}[!htbp]
    \centering
    \includegraphics[width=.975\linewidth]{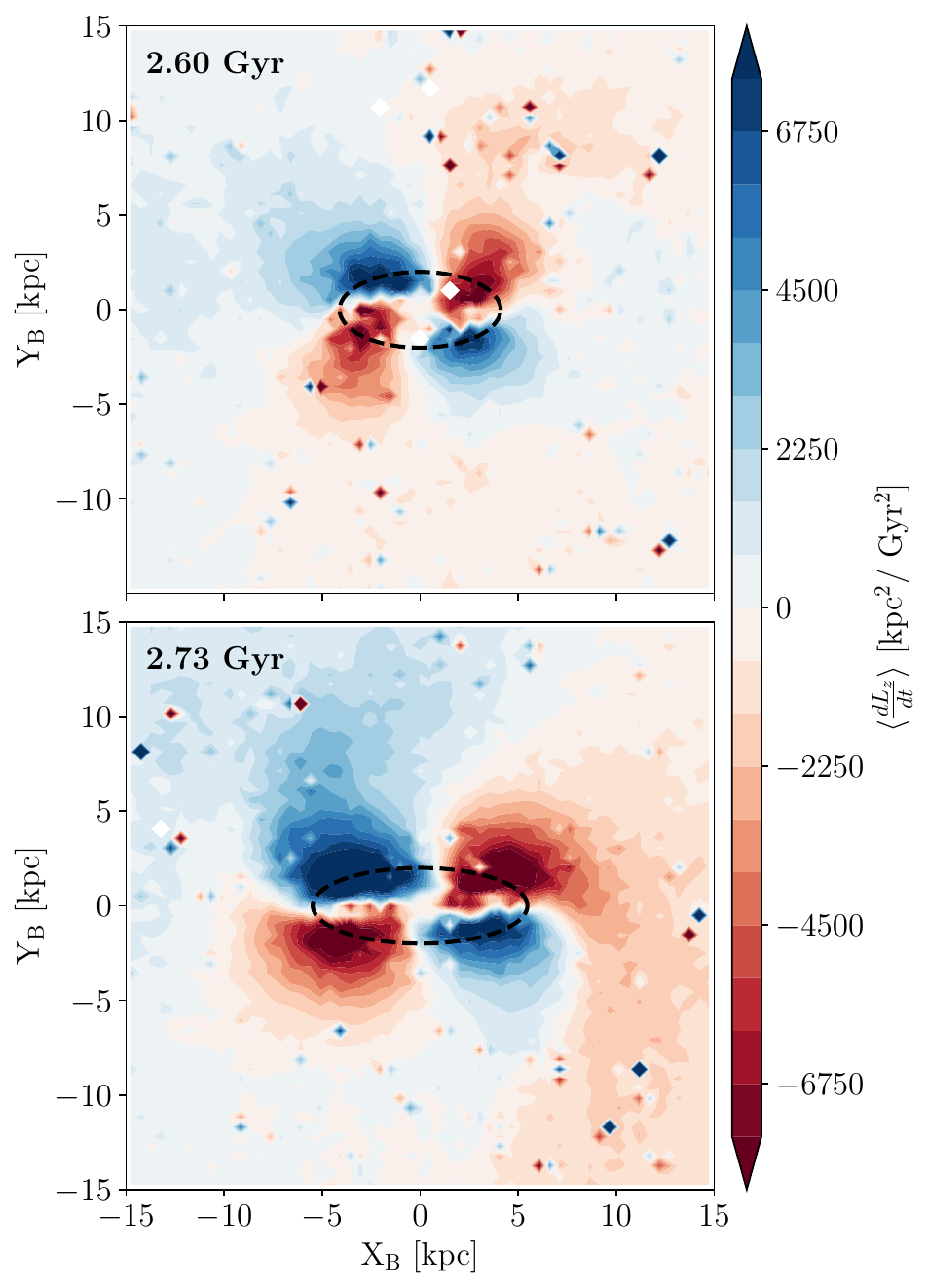}
    \caption{Mean vertical torque face-on maps $\langle\frac{dL_z}{dt}\rangle$, for two different times. The maps are oriented along the major and minor bar axes $\rm X_B$ and $\rm Y_B$. The bar region is shown in dashed ellipses, where the width along the $\rm X_B$ axis is the bar length at that time.}
    \label{fig:torque_maps}
\end{figure}

The time evolution of gas mass flux is shown in Fig.~\ref{fig:mass_flux}, through annuli at radii 1, 2 and 3~kpc. The highest mass influx occurs, predictably, at bar formation, while the reconnection events lead to positive mass influx. Outside of these events, the mass fluxes hover around net 0~$\rm M_\odot/yr$. The peaks of mass influx, similarly to the peaks of gas infall measured in the second panel of Fig.~\ref{fig:overdensity_t_r}, correlate with a slight delay with peaks of overdensities (first panel of Fig.~\ref{fig:overdensity_t_r}). The time delay between the peaks of mass influx at different radii reflects the infall time from more outer regions to more inner ones.

\begin{figure}[!htbp]
    \centering
    \includegraphics[width=\linewidth]{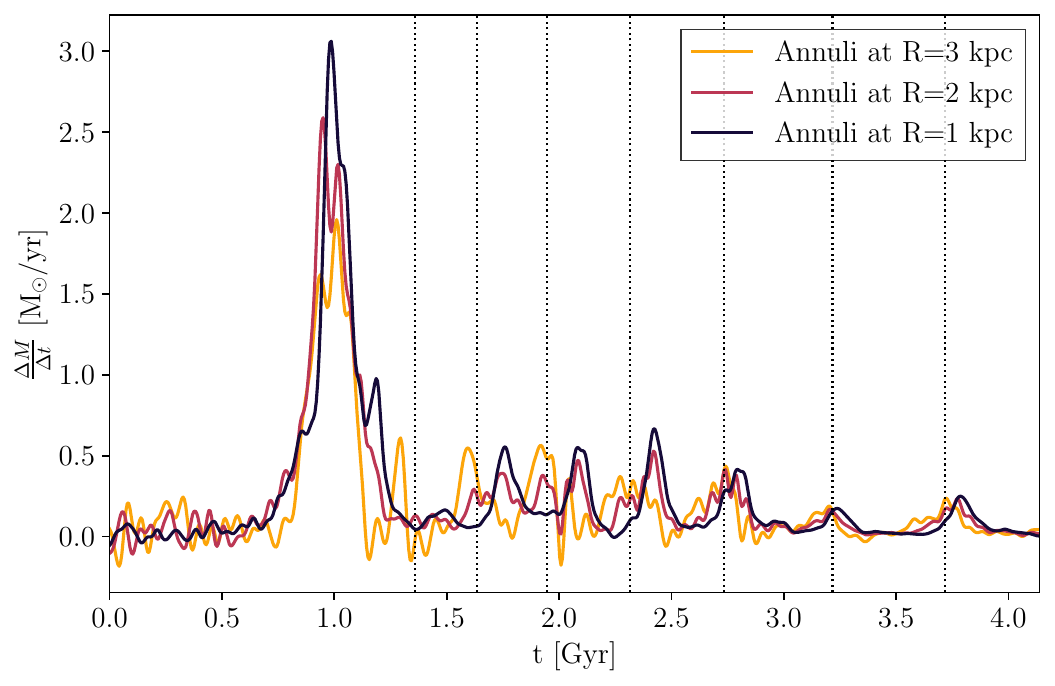}
    \caption{Time evolution of net gas mass flux through annuli at radius 1, 2 and 3~kpc. Light smoothing has been applied to mitigate noise from  the motion of gas. Peaks of overdensity are shown in dotted vertical lines. Positive values translate to mass inflows.}
    \label{fig:mass_flux}
\end{figure}

\section{$v/\sigma$ of the NSC and NSD}\label{app:v_over_sigma}

In Fig.~\ref{fig:v_over_sigma}, we plot the azimuthal velocity $v_\phi$, total velocity dispersion $\sigma$ and ratio $v_\phi/\sigma$ as a function of radius, for new stars within $|z| < .2$ kpc. The azimuthal velocity profile vanishes at R$\sim$0~pc, while it peaks at R=380~pc. The velocity dispersion profile is mostly flat within the NSC region, with a value of 65~km/s, followed by a rise and fall to a minimal value of 57~km/s at R=380~pc, and rises again at higher radii. This results in a $v/\sigma$ peaking at 2.25 at R=380~pc. Thus, the NSC region is mostly pressure supported, with a low azimuthal velocity but high dispersion, as a cluster-like structure would, while the NSD  region is rotationally supported, as a dynamically cool disc-like structure would, and as observations of these regions in the MW and external galaxies show.

\begin{figure}[!htbp]
    \centering
    \includegraphics[width=\linewidth]{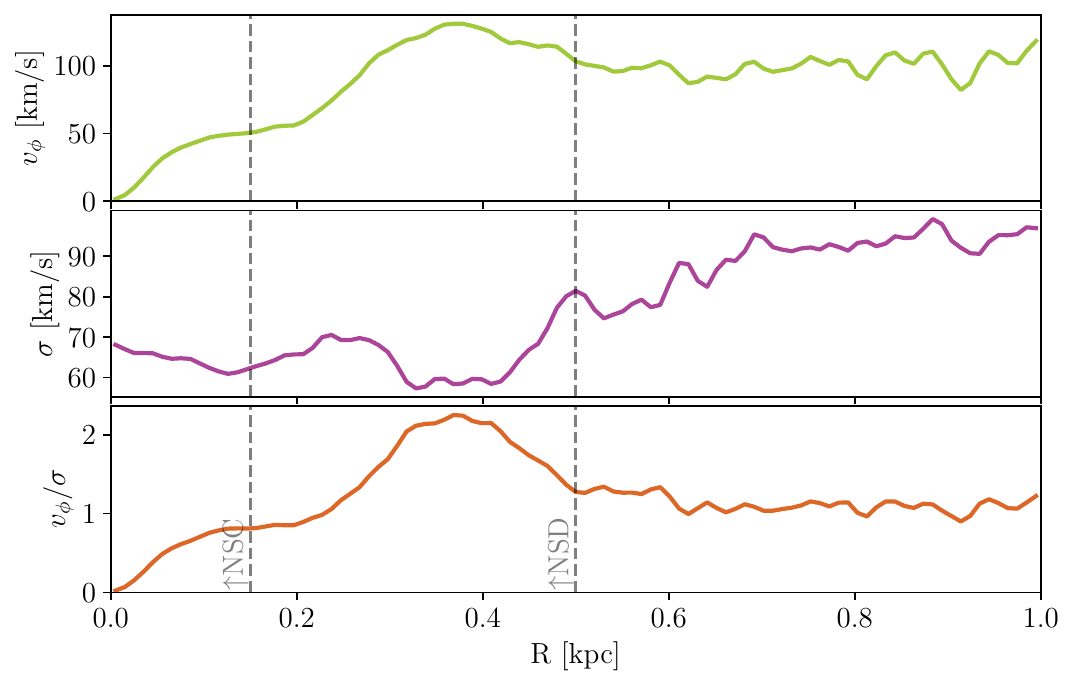}
    \caption{\textit{Top}: Azimuthal velocity profile $v_\phi(R)$. Middle: Total velocity dispersion profile $\sigma(R)$. \textit{Bottom}: Ratio $v_\phi/\sigma$ as a function of radius. The NSC and NSD regions defined previously are marked in dashed lines. Light smoothing has been applied.}
    \label{fig:v_over_sigma}
\end{figure}

\section{Spectrogram and spiral arms pattern speed estimation}\label{app:spectrogram}

In Fig.~\ref{fig:spectrogram}, we show a spectrogram of the simulation at t=1.58~Gyr, built from the time evolution of the m=2 mode of the radial stellar density distribution, i.e. by computing $\rm A_2(R) =\frac{\sum_i^{N_*}m_i\left(\cos(2\phi_i)+i\sin(2\phi_i)\right)}{\sum_i^{N_*}m_i}$ in different bins. To compute the frequency spectrum of $\rm A_2$, a Hanning function of width 1~Gyr is used centred at each timestep. Superimposed are the different resonances radial profiles (CR ($\Omega=\Omega_b$), ILR ($\Omega=\Omega_b+\frac{\kappa}{2}$) and OLR ($\Omega=\Omega_b-\frac{\kappa}{2}$) ).

The bar pattern speed at t=1.58~Gyr, as measured from the method described in \ref{sec:bar_bulge_spiral}, has a value of $\Omega_b=$31.2$\pm 0.2$~km/s/kpc. As expected, it coincides with a high power around this value in the bar region around 5~kpc. In a fashion similar to an idea first proposed by \citep{Oort_74,Wyse_89,Portinari_99} and further used in \cite{Mishurov_02} linking SFR to the difference between the rotation rate of the gas and the spiral arms pattern speed, we could compute the expected pattern speed difference between the bar and spiral arms by measuring the periods of the SFR bursts in the NSD and NSC region, as they should correspond to the 'beat' frequencies associated with $\Omega_b$ and $\Omega_{s}$. The relative frequency is thus given by 

\begin{equation}\label{eq:delta_omega}
\Delta\omega=\Omega_b-\Omega_{s}=\frac{2\pi}{2\rm T}{\,}
,\end{equation}
where the extra two factor in the denominator accounts for the m=2 modes. For t=1.58~Gyr, we measure a period of T=0.29~Gyr, leading to an expected $\Omega_s(\rm t=1.58~Gyr)=18.7\pm 5.5~\rm km/s/kpc$. This indeed coincides with a high power at radii outside the bar region $>5~\rm kpc$, where the spiral arms lie.

To quantify the applicability of this method, we measure the bar pattern speed and spiral arms pattern speed for t$>1.5~$Gyr using spectrograms. We estimate $\Omega_b$, $\Omega_s$ and their associated uncertainties by fitting Gaussian curves to the spectrogram signal within R=15~kpc. The spiral arms pattern speed computed in this way is shown in Fig.~\ref{fig:delta_omega} in blue, and for t=1.58~Gyr in Fig.~\ref{fig:spectrogram} shown by the black solid line. The periods of the SFR are measured from the time differences of its minima, computed with bins of width $\delta t=70~$Myr. Using Eq.~\ref{eq:delta_omega}, we compute the uncertainty $\delta\Delta\omega$ as $\delta\Delta\omega=\left|\frac{-2\pi}{2T^2}\delta t\right|$. The spiral arms pattern speed is then estimated by computing $\Omega_b-\Delta\omega$. In order to get an estimate of $\Omega_s$ that is independent of the pattern speeds measured from spectrograms, we use $\Omega_b$ as computed from the method of \cite{Dehnen_23} and shown in the bottom panel of Fig.~\ref{fig:bar_properties}, smoothed with a 1~Gyr Gaussian kernel, similarly to how spectrograms are built. The uncertainty on this estimate is then $\delta\Omega_s=\sqrt{\left(\delta\Delta\omega\right)^2+\delta\Omega_b^2}$. The resulting spiral arms pattern speed estimate is shown in Fig.~\ref{fig:delta_omega} in red. The two measurements display a similar trend of decreasing speed from $\Omega_s\sim20~$km/s/kpc at t=1.5~Gyr to $\Omega_s\lesssim 15~$km/s/kpc at t=4~Gyr, both in agreement within their uncertainty.

In this manner, one can retrieve the relative frequencies of the bar and spiral arms, just from the observed SFH of the NSD and NSC. If one has constraints on the bar pattern speed evolution and current value, as well as knowledge of the multiplicity of spiral arms, this method opens up a way to constrain the spiral arms pattern speed temporal evolution. While current SFR measurements do not reach the time resolution needed to reliably reconstruct the spiral arms pattern speed history, upcoming surveys such as MOONS \citep{Cirasuolo_20} will deliver much more refined data of the nuclear regions.

\begin{figure}[!htbp]
    \centering
    \includegraphics[width=\linewidth]{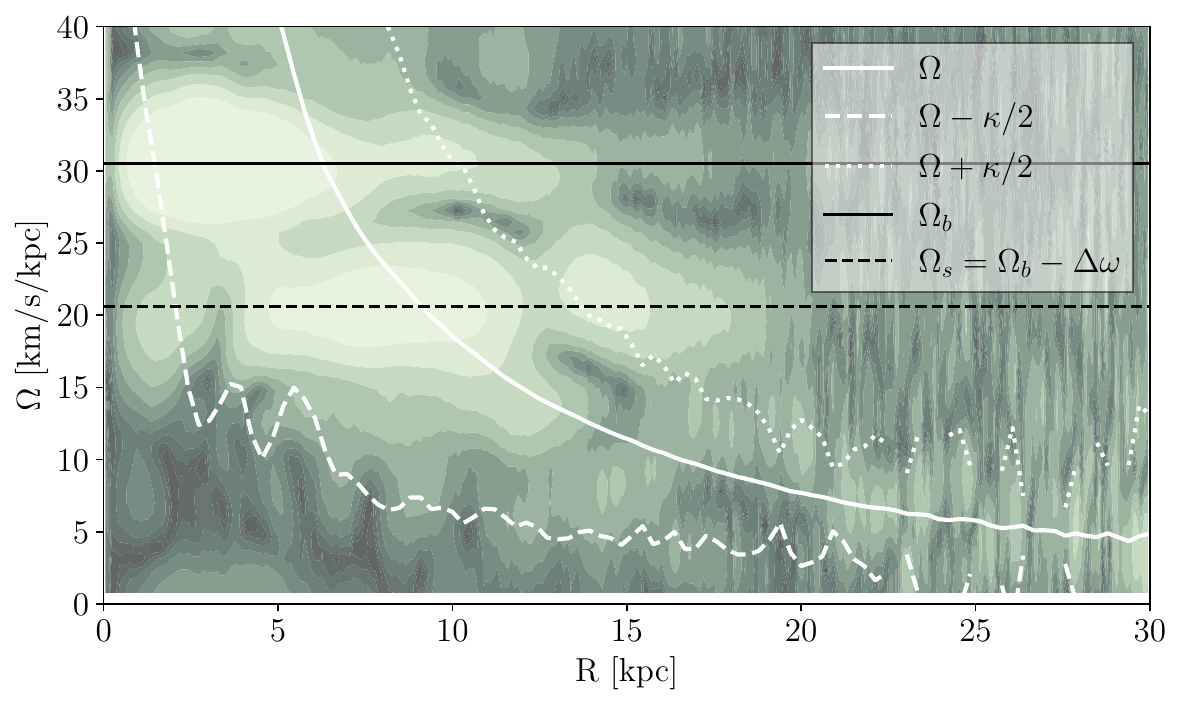}
    \caption{Spectrogram at t=1.58~Gyr, with a 1~Gyr window. The CR ($\Omega=\Omega_b$), ILR ($\Omega=\Omega_b+\frac{\kappa}{2}$) and OLR ($\Omega=\Omega_b-\frac{\kappa}{2}$) radial profiles, computed from the frozen potential using the epicyclic approximation, are superimposed in white respectively solid, dashed and dotted curves. The bar pattern speed $\Omega_b$ is shown as a solid black horizontal line and the
    spiral arms pattern speed $\Omega_{s}$ is shown as a dashed black horizontal line.}
    \label{fig:spectrogram}
\end{figure}

\begin{figure}[!htbp]
    \centering
    \includegraphics[width=\linewidth]{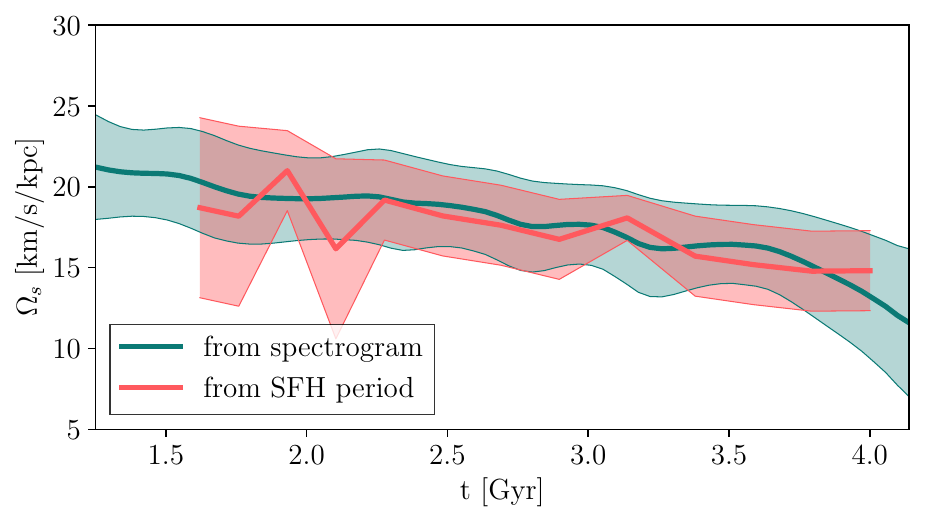}
    \caption{Time evolution of the spiral arms pattern speed, as measured from spectrograms with a 1~Gyr time window (blue) and from the periodic signal of the NSD SFH (red). Shaded areas show $\pm 1\sigma$ uncertainties.}
    \label{fig:delta_omega}
\end{figure}

\end{appendix}

\end{document}